\documentclass[amsmath,amssymb,11pt]{article}
\usepackage{jheppub2}
\pdfoutput=1
\usepackage{graphicx,epsfig}
\usepackage{float}
\usepackage{amsmath}
\usepackage{subfloat}
\usepackage[utf8]{inputenc}
\usepackage{hyperref}
\usepackage{caption}
\usepackage{mathrsfs}
\usepackage{subcaption} 
\usepackage{color}
\usepackage{overpic}
\usepackage[dvipsnames]{xcolor}
\usepackage{physics}
\usepackage[multiple]{footmisc}
\usepackage{comment}
\usepackage{tensor}
\usepackage[normalem]{ulem}
\usepackage{orcidlink}

\usepackage{pifont}

\newcommand{\beq}{\begin{equation}}
\newcommand{\eeq}{\end{equation}}

\newcommand{\be}{\begin{equation}}
\newcommand{\ee}{\end{equation}}
\newcommand{\ba}{\begin{eqnarray}}
\newcommand{\ea}{\end{eqnarray}}

\newcommand{\beqa}{\begin{eqnarray}}
\newcommand{\eeqa}{\end{eqnarray}}

\usepackage{tikz}
\usetikzlibrary{positioning}
\usetikzlibrary{intersections}
\usetikzlibrary{fadings} 
\usetikzlibrary{arrows.meta} 
\usetikzlibrary{arrows}

\tikzfading[name=fade out,
inner color=transparent!0,
outer color=transparent!100]

\definecolor{cherryblossompink}{rgb}{1.0, 0.72, 0.77}
\definecolor{lightblue}{rgb}{0.68, 0.85, 0.9}

\usetikzlibrary{decorations.pathmorphing}
\usetikzlibrary{decorations.pathreplacing,decorations.markings}

\usetikzlibrary{backgrounds,automata}

\begin{document}

\title{Rotating Extremal Black Holes in Einstein--Born--Infeld Theory}

\author[a]{Tom\'a\v{s} Hale,}
\emailAdd{tomas.hale@mff.utf.cuni.cz}
\affiliation[a]{Institute of Theoretical Physics Faculty of Mathematics and Physics,
Charles University V Holešovičkách 2, 180 00 Prague 8, Czech Republic}

\author[b]{Robie A.~Hennigar,}
\affiliation[b]{Centre for Particle Theory, Department of Mathematical Sciences, Durham University, Durham DH1 3LE, UK}
\emailAdd{robie.a.hennigar@durham.ac.uk}

\author[a]{David Kubiz\v{n}\'ak}
\emailAdd{david.kubiznak@matfyz.cuni.cz}

\abstract{We construct exact solutions that describe the near horizon region of extremal rotating black holes in Einstein--Born--Infeld theory.  Using generalized Komar integrals, we extract the electric charge and angular momentum from the near horizon geometries and study their deviations from the Kerr--Newman solution. We identify two features that are direct consequences of the nonlinearities of Born-Infeld theory. First, we find solutions which have vanishing charge but nontrivial electric and magnetic fields. Second, we find that extremal rotating black holes do not exist for sufficiently small charge and angular momentum. Based on analogy with the static black holes, we argue that it would be particularly interesting to construct the full rotating solutions in these parameter regions as they may provide examples of rotating black holes without Cauchy horizons.
}

\maketitle

\section{Introduction}

{\em Black holes} are among the most striking predictions of general relativity. With their strong gravitational fields, they provide a testbed for observing potential deviations from Einstein’s theory. 
Black holes also present 
profound conceptual challenges to our understanding of the world. These are especially acute in black hole interiors, where singularities and Cauchy horizons call into question the notion of predictivity in the laws of physics.  Fully addressing these challenges will require going beyond general relativity in some way. 



Although we shall not address it directly, one of the main inspirations for our work here is the question of {\em strong cosmic censorship}. This conjecture attempts to formalize the physical notion that Cauchy horizons should not arise in realistic solutions of the Einstein equations. However, the familiar Kerr--Newman family of metrics \textit{do} contain Cauchy horizons, and consistency with the conjecture then requires that such Cauchy horizons are non-generic. There are very good reasons to believe that general relativity itself contains  the cure for this disease. Namely, Cauchy horizons are prone to mass inflation instabilities the backreaction of which can render the Cauchy horizon singular~\cite{Poisson:1989zz}. In this sense, Cauchy horizons would be an artefact of \textit{fine-tuning}: present only under the unrealistic assumption of an exact symmetry.

However, strictly speaking this may not be true -- or at least more subtle. 
For example, in the case of charged de Sitter black holes, the classical instabilities associated with the Cauchy horizon are insufficient to render it singular~\cite{Cardoso:2017soq, Dias:2018etb}. In these cases, one needs to appeal to a different mechanism to enforce predictivity. One such mechanism is to regard these objects not only as requiring fine-tuning in the space of metric deformations, but also fine-tuning in the \textit{space of theories}. For example, if one considers the physically very sensible modification of classical general relativity to \textit{semi}-classical gravity, then one finds strong semiclassical instabilities for inner horizons, stronger even than their classical counterparts~\cite{Hollands:2019whz}. Along similar lines, various recent works have found that the Cauchy horizon of the Reissner-Nordstr{\"o}m black hole (more specifically, its AdS generalization) is eliminated under very general circumstances upon coupling Einstein--Maxwell theory to additional charged scalars, or including mass terms for the gauge fields~\cite{Hartnoll:2020rwq, Cai:2020wrp, An:2021plu, DeClerck:2023fax}. It therefore is an interesting question to consider just how general are Cauchy horizons in the {\em space of deformations} of Einstein's gravity and of the matter content,
even in the presence of significant symmetry. 


In this work we preserve Einstein's {\em general relativity} and couple it to
{\em nonlinear electrodynamics (NLE)}, a deformation of Maxwell theory in which the electrodynamic Lagrangian is nonlinear. The prototypical example of an NLE is the Born--Infeld theory, which was originally proposed to tame the infinite self-energy of point charges~\cite{Born:1934gh}. Among its notable features are: a finite self-energy for point sources; causal, hyperbolic equations of motion; absence of birefringence; exact electromagnetic duality even at the nonlinear level; and a stress tensor that satisfies the dominant and strong energy conditions~\cite{Plebanski:1970zz,Bialynicki-Birula:1992rcm,Gibbons:1995cv,Russo:2024xnh}. The theory gained renewed significance in the 1980s when it was realized to arise as the low-energy effective action for gauge fields and D-branes in string theory~\cite{Fradkin:1985qd,Leigh:1989jq}.

In a recent paper~\cite{Hale:2025ezt}, we proved that in any theory of NLE which is (i) causal, and (ii) predicts finite self-energy for point charges, the Cauchy horizon of the spherically symmetric Reissner--Nordstr{\"o}m metric is \textit{eliminated} for {\em weakly charged} black holes. In particular, this is the case of the Einstein--Born--Infeld black holes, for which, in addition, there exists a {\em universal bound} on charge below which no Cauchy horizon can exist (black holes below this bound necessarily feature a single horizon with a spacelike singularity). The requirement of being \textit{weakly charged} may at first seem odd, but this makes perfect physical sense. The effects of NLEs is most important when the electromagnetic field strength is large. When a Reissner--Nordstr{\"o}m black hole is weakly charged, its Cauchy horizon occurs at short distances and hence large field strengths. The main limitation of~\cite{Hale:2025ezt} is the requirement of spherical symmetry, as it says nothing about the fate of Cauchy horizon in the {\em rotating} case. Could the Cauchy horizon of a (weakly charged) Kerr--Newman solution be also eliminated after deforming the Maxwell theory to the appropriate NLE one?  It is the purpose of this work to make progress on this issue. 

In fact, apart from three dimensions (see Appendix~\ref{App:BTZ} for a lightning review) very little is known about rotating black holes in NLEs whatsoever, making this an interesting question in its own right. What is known comes largely from higher-dimensional perturbative studies~\cite{Allahverdizadeh:2013oha, Allaverdizadeh:2013rua}, slow rotation expansions~\cite{Kubiznak:2022vft}, or from constructing rotating black branes, which are related to the static solutions by a simple boost~\cite{Hendi:2010kv,Hendi:2010zz}.\footnote{Note that there is a vast literature on rotating black holes in NLEs generated by a (generalized) Newman--Janis formalism from static solutions. However, the obtained spacetimes do not satisfy the equations of motion, unless some additional (ad hoc) energy momentum  tensor is introduced. } While interesting for their own reasons, none of these methods tell us a great deal about the strong field properties of rotating black holes in NLEs. Only very recently has the issue of four-dimensional rotating black holes in NLEs received serious attention. In~\cite{Cheng:2025kfz}, the Einstein--Born--Infeld equations were solved numerically, obtaining a generalization of the Kerr--Newman solution. This allowed the authors to study for the first time fully nonperturbative aspects of geodesic motion, thermodynamics, and the gyromagnetic ratio of rotating Born--Infeld black holes. Moreover, based on the behaviour of the temperature, the results of~\cite{Cheng:2025kfz} appear to be consistent with the existence of
single horizon rotating Born--Infeld black holes in some parameter region.
Unfortunately, those results break down near extremality and do not allow for the interior solution to be assessed due to the type of numerical methods employed, hence the existence of single horizon solutions cannot be unambiguously established.

The goal of our work is in some sense more modest. Instead of considering full rotating solutions, we focus on the {\em near horizon region} of an extremal rotating black hole. Namely, 
we will obtain for the first time an {\em exact analytic} solution to the Einstein--Born--Infeld equations that describes the near horizon region of an extremal rotating black hole, complementing the work in \cite{Cheng:2025kfz}. Using generalized Komar integrals, we will be able to extract the electric charge and angular momentum from the near  horizon geometry and study their relationship to the horizon area. More importantly, we will be able to show that there are regions in the parameter space where no such near horizon geometries exist. It is precisely in these regions that in the corresponding static limit the Born--Infeld black holes possess only a single horizon and do not admit extremal solutions.  We regard this results 
as highly suggestive of the existence of rotating black holes without Cauchy horizons.

Our paper is organized as follows. The basic ingredients for our construction, including a review of theories of NLE, and a description of near horizon geometries and their physical characteristics, are gathered in Sec.~\ref{Sec:preliminaries}. Sec.~\ref{Sec:3} summarizes the properties of spherically symmetric Einstein--Born--Infeld black holes; the universal charge bound on the existence of Cauchy horizons is overviewed and shown to directly follow  from the requirement on the existence of the corresponding near horizon geometry. To set the stage for our study of near horizon rotating geometries, we revisit the Kerr--Newman solution in Sec.~\ref{Sec:4}. 
The main section of our study is Sec.~\ref{Sec:5}, where 
the novel results regarding the near horizon geometries of extremal rotating Einstein--Born--Infeld black holes are presented.
We conclude in Sec.~\ref{Sec:6}. Appendices \ref{App:A} and \ref{app:numerical_check} contain additional technical results and summarize numerical verification of our analytic solution. For comparison, we have also added Appendices~\ref{App:BTZ} and 
\ref{App:strings} where we review the rotating Einstein--Born--Infeld 
BTZ black hole and
black string `cousins' of spherical black holes studied in the main text. In both cases (although it has not been appreciated in the literature), the Born--Infeld electrodynamics eliminates the Cauchy horizon in certain regions of the parameter space, adding qualitative support to our argument.

\section{Preliminaries}
\label{Sec:preliminaries}

In this section, we gather together various ingredients we shall need later. This includes the 
ans{\"a}tze for the extremal near horizon geometries, the theories and their equations of motion, as well as Komar-type integrals that allow one to extract physical quantities from the near horizon metrics.

\subsection{Extremal near horizon metrics and their regularity}

Quite a lot of attention has been paid to the near horizon geometries of extremal black holes over the last couple of decades. This is largely due to the fact that in the near horizon region additional symmetries emerge which simplify various problems of interest -- see, e.g.,~\cite{Kunduri:2007vf,Kunduri:2013gce}. Here we are primarily interested in two classes of extremal near horizon geometries: the four-dimensional static and rotating near horizon geometries of topologically spherical black holes.

Under the restriction to {\em static} metrics, the most general extremal near horizon geometry and gauge field read
\begin{align} \label{eq:ads2xs2}
{\rm d}s^2 &= L_1^2 \left(-r^2 {\rm d}t^2 + \frac{{\rm d}r^2}{r^2}\right) + L_2^2 \left(\frac{{\rm d} y^2}{1-y^2}  + (1-y^2) {\rm d} \phi^2\right) \, ,
\nonumber\\
A &= q r {\rm dt} \, .
\end{align}
In this metric $L_1, L_2$ and $q$ are constants that must be determined by the equations of motion of the corresponding theory. 
$y$ is a compact coordinate $y \in [-1,1]$ with $\pm 1$ corresponding to the poles of the sphere, and the azimuthal angle $\phi$ is periodic with period $2 \pi$.

In the case of {\em rotating} near horizon geometries, the situation is somewhat more complicated with the metric and gauge field now reading~\cite{Astefanesei:2006dd,Cano:2019ozf}\footnote{Strictly speaking, we should be including here an arbitrary function $p(x)$ multiplying the AdS$_2$ sector of the metric, i.e. $p(x) {\rm d} s^2_{{\rm AdS}_2}$. However, it is straightforward to show that the difference between the $xx$ and $\phi\phi$ components of the field equations, i.e {$G_x^x - G_\phi^\phi = 8 \pi (T_x^x - T_\phi^\phi)$} implies that $p(x) = (x^2 + a^2)$ is the most general structure that is (i) $x \to -x$ symmetric, and (ii) consistent with the field equations for Einstein--Born--Infeld theory. Thus, there is no loss of generality here.}
\begin{align}
    {\rm d}s^2 &= \left(x^2 + n^2 \right) \left(-r^2 {\rm d}t^2 + \frac{{\rm d}r^2}{r^2}\right) + \frac{{\rm d} x^2}{f(x)} + N(x)^2 f(x) \bigl({\rm d}\psi - 2 n r {\rm d} t \bigr)^2\,,\nonumber
    \\
    A &= h(x) \bigl({\rm d}\psi - 2 n r {\rm d} t \bigr) \, .
\end{align}
The fields depend on three functions of one variable, $h(x), N(x)$ and $f(x)$  and a constant parameter $n$. For our cases, it is possible to make further simplifications and cast the metric into a more convenient form. 

We are interested in Einstein gravity coupled to theories of nonlinear electrodynamics. In this case, it is not difficult to show that the general solution of the Einstein-NLE equations require that $N(x)$ is a constant, which we can set to unity without loss of generality.  The zeroes of the function $f(x)$ are poles of the sphere and define the range of the coordinate~$x$. Here we are interested only in the case where $f(x)$ is an even function -- one can show that if $f(x)$ is odd, then the metric contains a NUT charge~\cite{Cano:2019ozf}. Therefore, if $x_0$ is the smallest positive zero of $f(x)$, then the coordinate range will be $x \in [-x_0, x_0]$. The absence of conical singularities at $\pm x_0$ requires that
\be 
\psi \sim \psi + \frac{2 \pi}{\omega}
\ee
where 
\be 
\omega \equiv - \frac{f'(x_0)}{2} \, .
\ee

Let us then introduce two new coordinates and redefine certain functions appearing in the metric:
\be 
x = x_0 y \, , \quad \psi = \frac{\phi}{\omega} \, , \quad g(y)  = \frac{f(y x_0)}{x_0^2} \, .
\ee
In terms of these new quantities, the metric and gauge field that we shall consider read
\begin{align}\label{eq:ext_geom}
{\rm d}s^2 &= \left( x_0^2 y^2 + n^2 \right) \left( -r^2 {\rm d}t^2 + \frac{{\rm d} r^2}{r^2} \right) + \frac{dy^2}{g(y)} + x_0^2g(y) \left( \frac{{\rm d}\phi}{\omega} - 2 n r {\rm d}t \right)^2 \, ,
\nonumber\\
A &= h(y) \Bigl(\frac{{\rm d} \phi}{\omega} - 2 n r {\rm d}t \Bigr) \, .
\end{align} 
Here $y$ is now a compact coordinate with range $y\in[-1, 1]$, $\phi$ is an azimuthal coordinate with periodicity $2 \pi$. The parameters $x_0$, $n$ and $\omega$ are constants.
Regularity of the geometry requires that
\be \label{eq:regularity}
g(\pm 1) = 0 \, , \quad g'(\pm 1) = \mp \frac{2 \omega}{x_0} \, .
\ee

On the other hand, the gauge field as written above is not regular at the poles $y = \pm 1$. This is not a problem for obtaining a solution of the electromagnetic equations, but it will be important when obtaining the angular momentum (see below), for which we must work in a regular gauge. In the case of vanishing magnetic charge, a simple transformation can be made to obtain a regular gauge field,
\be 
A_{\rm reg}= h(y)\left(\frac{{\rm d} \phi}{\omega} - 2 n r {\rm d}t \right) - h(1)\frac{{\rm d} \phi}{\omega} \, .
\ee

Before concluding this subsection, let 
summarize the parameter counting for the rotating near horizon geometries. 
The metric and gauge field solutions \eqref{eq:ext_geom} are described by 6 parameters in total.
Namely, 
$\{n,x_0,\omega\}$ appear directly in \eqref{eq:ext_geom}. At the same time,  the equations of motion comprise of a second order equation for $h(y)$ and a first order equation (see below) for $g(y)$ which together give 3 integration constants, e.g.: \be 
\{h_0\equiv h(0)\,,\  h_1\equiv \frac{\dd h}{\dd y}(0)\,,\  g_1\equiv\frac{\dd g}{\dd y}(0)\}\,.
\ee
Demanding the absence of NUT charge makes $g(y)$ an even function, fixing $g_1=0$ (smooth horizon across $y=0$).
The absence of a magnetic charge makes $h(y)$ an even function and fixes $h_1=0$  (see below).
$x_0$ and $\omega$ are `convenience parameters', which fix the parameter ranges of $y \in \left[-1,1 \right]$ and $\phi\in[0,2\pi)$. The value of $\omega$ is determined by the regularity condition 
$g'(\pm1)=\mp\frac{2\omega}{x_0}\,.$
The spherical horizon topology condition $g(\pm1)=0$ now provides one more relation between the parameters $\{n,x_0,h_0\}$, which correspond to the physical quantities of area, angular momentum and charge. Therefore, the solution space can be fully described by {\em two} independent parameters, as is to be expected for a charged and rotating extremal black hole.

\subsection{Equations of motion: Einstein-NLE theories}

In this work, we will study the near horizon geometries just described as solutions to Einstein gravity coupled to theories of electrodynamics. We will consider only Einstein--Maxwell and Einstein--Born--Infeld theories, but it is straightforward to present the general formalism for any theory of electrodynamics within the Plebanski class \cite{plebanski1970lectures}. 

Consider a (parity even) theory of electromagnetism with Lagrangian ${\cal L}$ built from the two basic electromagnetic invariants,
\be 
\mathcal{L} = \mathcal{L}\left(\mathcal{S}, \mathcal{P}^2\right) \, , \quad \mathcal{S} = \frac{1}{2} F_{\mu\nu} F^{\mu\nu} \, ,\quad \mathcal{P}=\frac{1}{2} F_{\mu\nu}(\star F)^{\mu\nu} \, .
\ee
Here $F_{\mu\nu} = ({\rm d} A)_{\mu\nu}$ is the electromagnetic field strength tensor, while $A_{\mu}$ is the vector potential. It is useful to define the object $D_{\mu\nu}$ according to
\be\label{eq:nle_D} 
D_{\mu\nu} \equiv -2 \frac{\partial \mathcal{L}}{\partial F^{\mu\nu}} = -2 \Bigl(\mathcal{L}_\mathcal{S} F_{\mu\nu} + \mathcal{L}_\mathcal{P} (\star F)_{\mu\nu} \Bigr) \, ,
\ee
where 
\be 
\mathcal{L}_\mathcal{S} = \frac{\partial \mathcal{L}}{\partial \mathcal{S}} \, , \quad \mathcal{L}_\mathcal{P} = \frac{\partial \mathcal{L}}{\partial \mathcal{P}} \, .
\ee
The Einstein-NLE equations then read
\be \label{eq:ein_nle}
{\rm d}F = 0 \, , \quad {\rm d} \star D = 0 \, , \quad G_{\mu\nu} = 8 \pi T_{\mu\nu} \, ,
\ee
where the electromagnetic stress-energy tensor is given by
\be 
T^{\mu\nu} = - \frac{1}{4 \pi } \left(2 F^{\mu \sigma}F^{\nu}{}_{\sigma} \mathcal{L}_\mathcal{S} + \mathcal{P} \mathcal{L}_\mathcal{P} g^{\mu\nu} - \mathcal{L} g^{\mu\nu} \right) \, .
\ee

In this work, we will be almost exclusively concerned with two theories of electromagnetism. The first is simply the Maxwell theory for which we have
\be 
\mathcal{L}_{\rm Max} = -\frac{1}{2} \mathcal{S} \, ,
\ee
and the second is the Born--Infeld theory~\cite{Born:1934gh}, which in four dimensions can be expressed as
\be 
\mathcal{L}_{\rm BI} = b^2 \left(1 - \sqrt{1 + \frac{\mathcal{S}}{b^2} - \frac{\mathcal{P}^2}{4 b^4}}\right) \, .
\ee
In the Born--Infeld action, the parameter $b$ gives the maximum allowable field strength. In geometric units, $[b] = (\text{length})^{-1}$. The Maxwell theory is recovered in the limit $b \to \infty$. 

\subsection{Physical properties}

It is well-known that {\em modified Komar} integrals may be used to compute the angular momentum and electric/magnetic charges of solutions from the near horizon geometries~\cite{Hanaki:2007mb, Kunduri:2013gce}. Here we present the generalization of those results to general NLEs. 

For a general NLE in the Plebanski class, the Gauss' law for the electric and magnetic charges take the form
\be 
Q_e =  \frac{1}{4 \pi} \int_{S} \star D \, , \qquad Q_m  = \frac{1}{4 \pi} \int_{S} F\,,
\ee
where the two-form $D_{\mu\nu}$ was defined above in eq.~\eqref{eq:nle_D}. The angular momentum is given by  
\be 
J = -\frac{1}{16 \pi } \int_S \left( \star {\rm d} m + 4 (m \cdot A) \star D \right) \, ,
\ee
where $m = m_\mu {\rm d}x^\mu$ is the rotational Killing field. These expressions allow one to obtain the electric/magnetic charges and angular momentum directly from the near horizon geometry. Note that to apply the Komar integral for the angular momentum, the gauge field should be regular and with $\mathscr{L}_m A = 0$ on the horizon. While the formula for the charge is well-known, the expression for the angular momentum is new. This extends the surface-independent form of the Einstein-Maxwell Komar integral to general theories of NLE.  As we will see later, these charges will satisfy a `near horizon' first law, which is a nontrivial consistency check on their validity.

Let us also note that the horizon area is easily computable from the near horizon geometry,
\be 
A = \frac{4 \pi x_0}{\omega} \, .
\ee
Unfortunately, it is not possible to extract the mass from the near horizon metric. That is because the transformations applied to the full spacetime metric to bring it to the near horizon form involve a singular rescaling of the time coordinate. However, being able to compute the electric/magnetic charges, angular momentum, and area will be sufficient for us to obtain direct relationships between the physical parameters whose validity is not restricted to the near horizon region.

\section{Static black holes in Born--Infeld theory}\label{Sec:3}
Let us start by reviewing the static spherically symmetric black holes in Born--Infeld theory. As mentioned in the introduction, there exists a universal bound on charge, below which no extremal solutions can exist. As we shall see, such a bound can be easily extracted from  considering the near horizon extremal geometry. 

\subsection{Schwarzschild-like black holes and universal bound}\label{sec:universal bound BI}

The spherically symmetric black holes in Einstein--Born--Infeld theory have been studied long time ago, e.g. \cite{GarciaD:1984xrg, deOliveira:1994in, Fernando:2003tz, Dey:2004yt, Cai:2004eh}.\footnote{
In fact, the so called `{\em BIonic solution}'', or, in terminology of the current paper, a marginal solution without a horizon, was already obtained in 1935 by Hoffmann \cite{Hoffmann:1935ty} by considering `a different choice' of boundary conditions.}
The solution takes the standard form 
\be 
\dd s^2=-f\dd t^2+\frac{\dd r^2}{f}+r^2 \dd\Omega_2^2\,,\quad A=-\phi \dd t\,,
\ee 
where $\dd\Omega_2^2=\sin^2\theta \dd \varphi^2+\dd \theta^2$ is the spherical metric element, and 
\ba 
f&=&1-\frac{2M}{r}+\frac{2b^2}{r}\int_r^\infty\Bigl(\sqrt{r^4+\frac{Q^2}{b^2}}-r^2\Bigr)\,\dd r\label{D2}\nonumber\\
&=&
1-\frac{2M}{r}+\frac{2b^2r^2}{3}\Bigl(1-\sqrt{1+\frac{Q^2}{b^2 r^4}}\Bigr)+\frac{4Q^2}{3r^2} {}_2F_1\Bigl(\frac{1}{4},\frac{1}{2}; \frac{5}{4};-\frac{Q^2}{b^2r^4}\Bigr)\,,\\
\phi&=&\frac{Q}{r} {}_2F_1\Bigl(\frac{1}{4},\frac{1}{2}; \frac{5}{4};-\frac{Q^2}{b^2r^4}\Bigr)\,,
\ea 
with ${}_2F_1$ being the hypergeometric function. Here, the parameter $M$ represents the gravitational mass, the parameter $Q$ is the asymptotic charge of the solution, and the field strength takes a simple form:
\be 
F=E \dd r\wedge \dd t\,,\quad E=-\phi'(r)=\frac{Q}{\sqrt{r^4+Q^2/b^2}}\,.
\ee

\begin{figure}[t]
    \centering
\includegraphics[width=0.45\linewidth]{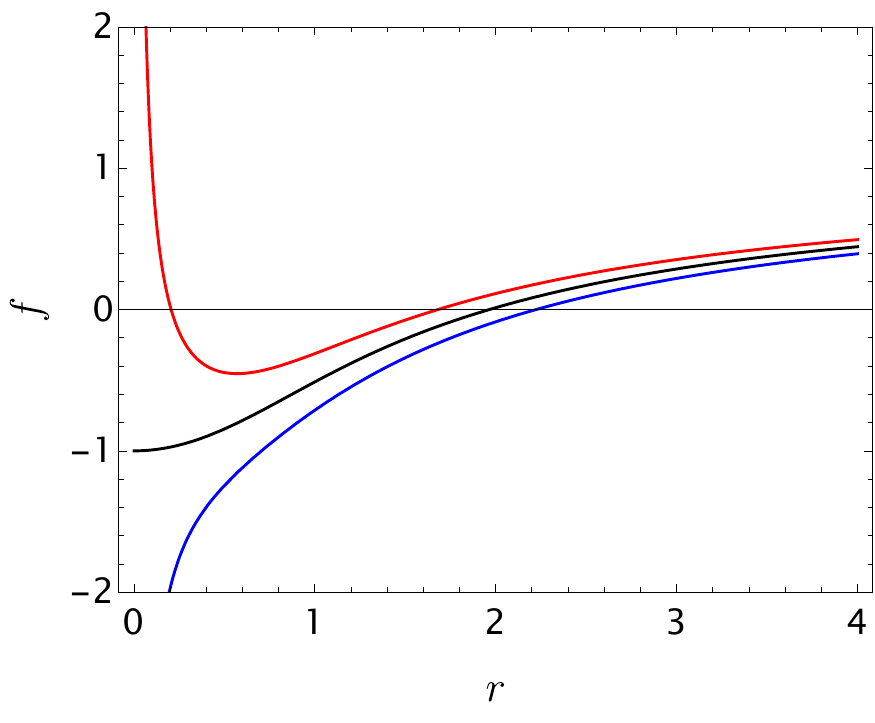}
\includegraphics[width=0.45\linewidth]{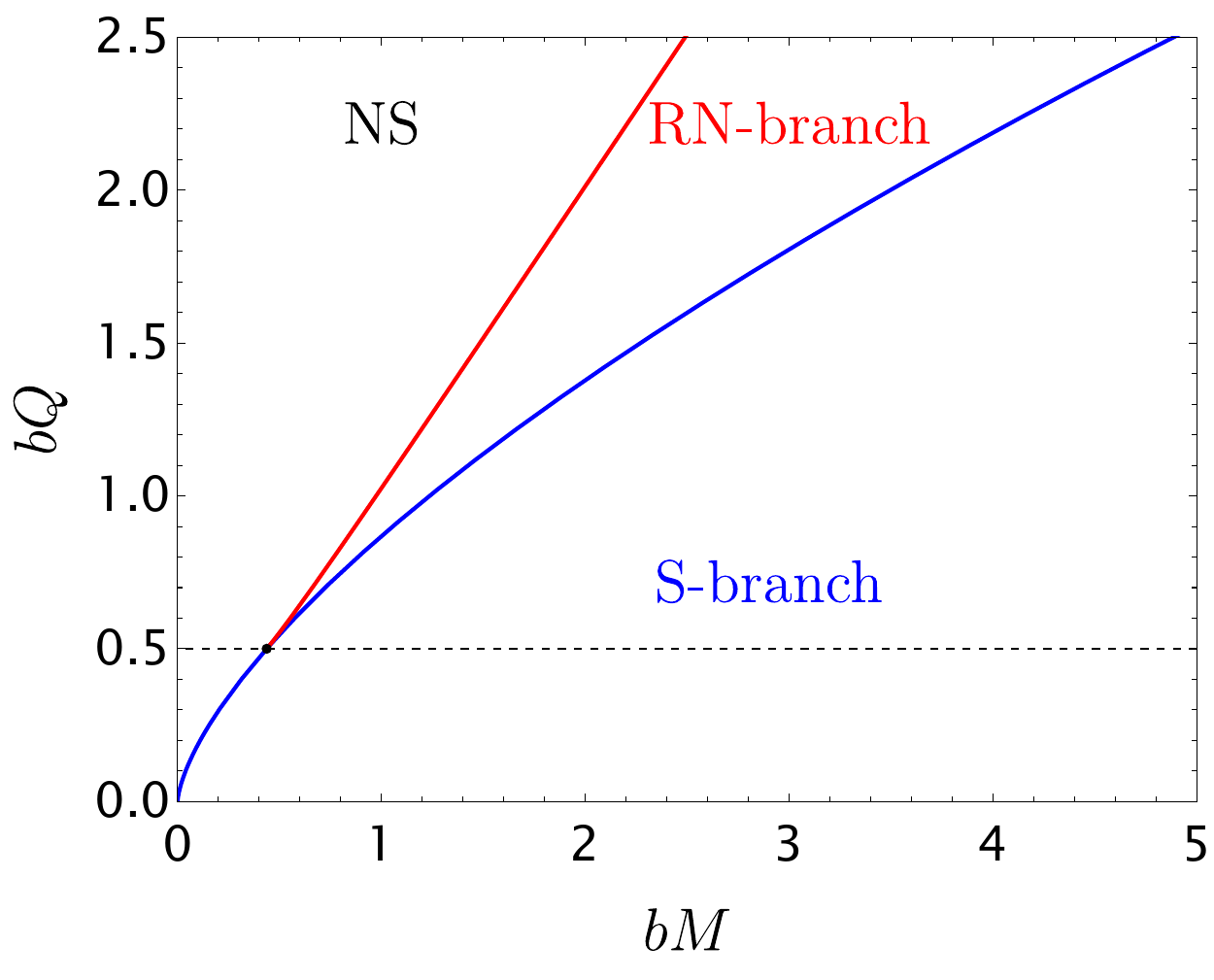}
    \caption{{\bf Static black holes in Born--Infeld theory}. {\em Left.} The behavior of metric function $f$ for the two possible types of static Born--Infeld black holes. The $M>U_{\mbox{\tiny self}}^{(0)}$ S-branch is illustrated by blue curve and the $M<U_{\mbox{\tiny self}}^{(0)}$ RN-branch is displayed by red curve. The two branches are separated by the  $M=U_{\mbox{\tiny self}}^{(0)}$ marginal case (black curve). {\em Right.} The $(bM, bQ)$ parameter space of static Born-Infeld black holes. We observe S-branch black holes  (below blue curve) and the RN-branch (between red and blue  
    curves). NS stands for naked singularity and the red
    curve corresponds to extremal black holes -- it terminates at the intersection with the (marginal) blue curve -- the `star point'. We observe a universal {\em charge gap} $|Q|=1/(2b)$ for the existence of extremal black holes (black dotted horizontal line). Note that there is a similar mass gap (not explicitly highlighted). Units in both diagrams were chosen so that $b=1$.  }
    \label{fig:BIbound}
\end{figure}

As is well known, the above solution describes two distinct branches of black hole solutions, see Fig.~\ref{fig:BIbound} (left). This can be seen from the expansion of $f$ around $r=0$. Namely, we find 
\be 
f=1-\frac{2(M-U_{\mbox{\tiny self}}^{(0)})}{r}-2b|Q|+O(r)\,,
\ee 
where 
\be 
U_{\mbox{\tiny self}}^{(0)}=-\frac{1}{4\pi}\int T^t_t\, \dd V=b^2\int_0^\infty \Bigl(\sqrt{r^4+\frac{Q^2}{b^2}-r^2}-r^2\Bigr)\,\dd r=\frac{1}{6}\sqrt{\frac{b}{\pi}}|Q|^{3/2}\Gamma\Bigl(\frac{1}{4}\Bigr)^2\,
\ee 
is the electrostatic self energy of the solution. Clearly, when $M>U_{\mbox{\tiny self}}^{(0)}$, the black hole possesses a spacelike singularity.
Moreover, it has a single horizon, ane because of its reminiscence to the Schwarzschild solution, the corresponding branch is called the (Schwarzschild-like) {\em S-branch}. On the other hand, when the self energy is smaller than the gravitational mass, $M<U_{\mbox{\tiny self}}^{(0)}$, the singulatrity is timelike and the behavior is Reissner--Nordstr{\"o}m-like, with two, one extremal, or no horizons. The corresponding branch of solutions is called the {\em RN-branch}.  

Interestingly, there is a {\em universal bound} for the existence of RN-branch, which imposes a lower bound on the mass and charge of such black holes, see Fig.~\ref{fig:BIbound} (right). This occurs, when the extremal black hole `curve' (displayed in red), obtained from 
\be 
f(r_E,M,Q,b)=0=f'(r_E,M,Q,b)\,,
\ee 
intersects the marginal line (blue curve) between the RN and S-branches, namely  
\be 
M=U_{\mbox{\tiny self}}^{(0)}\,.
\ee 
For Born--Infeld black holes 
this happens at a special `point'
\be 
M=M_\star=\frac{\Gamma(1/4)^2}{12\sqrt{2\pi}b}\,,\quad |Q|=Q_\star=\frac{1}{2b}\,.
\ee 
One can easily show that this `star solution' has vanishing radius, $r_E=0$, and precisely corresponds to `extreme black points' recently studied in \cite{Sokolov:2025ara, Sokolov:2025vtl}. Most importantly for us, and since the RN-branch black holes cannot exist for 
\be \label{gap}
|Q|<Q_\star=\frac{1}{2b}\,,
\ee 
there are {\em no extremal} black holes in this range -- there is a {\em charge} (and also mass) {\em gap} for the existence of static extremal black holes in the Born--Infeld theory. It also means that, if any black holes are found in this range, they must necessarily be S-type and thence must feature no Cauchy horizon. As we shall see next, the existence of the above bound can also be obtained from the analysis of the near horizon extremal black hole geometry.
Note also that the universal bound is absent for the rotating black strings reviewed in Appendix~\ref{App:strings}.

\subsection{Extracting the bound from near horizon geometry}

Let us now show that the universal bound on the charge can also be obtained directly from the near-horizon geometry \eqref{eq:ads2xs2}. Substituting these ans{\"a}tze  into the Einstein--Born--Infeld equations constrains the radii of the AdS$_2$ and S$_2$ factors as follows:
\be 
L_1 = \frac{\sqrt{4b^2q^2+1}}{2b} \, , \quad L_2 = \frac{\sqrt{4b^2q^2-1}}{2b}\,,
\ee
where the parameter $q$ is easily found to be related to the electric charge:
\be 
Q = \frac{1}{4 \pi } \int_S \star D = q \, .
\ee
We then see directly that there is a minimum charge required for the existence of extremal near-horizon geometries, 
\be 
Q \ge Q_{\rm min} = \frac{1}{2 b} \, ,
\ee
in accordance with the above gap \eqref{gap}.
When $Q = Q_{\rm min}$, the sphere radius vanishes, $L_2 = 0$, and the solution exhibits a null singularity at the location of the horizon. As we will see below, analogous constraints exist on the charge and angular momentum of extremal rotating black holes in Einstein--Born--Infeld theory.

Note also that in the Maxwell limit, $b \to \infty$,  the curvature radii of the sphere and AdS$_2$ are the same, 
\be 
\lim_{b \to \infty} L_1 = \lim_{b \to \infty} L_2  = Q \, .
\ee
Thus, we reproduce the well-studied near horizon geometry of the extremal Reissner--Nordstr{\"o}m solution.

Finally, it is easy to calculate the horizon area of the near horizon geometry, it reads
\be\label{Area:static} 
A=4\pi L_2^2=4\pi Q^2-\frac{\pi}{b^2}\,.
\ee 
The non-linear modification thus gives rise to a constant shift in the entropy of the extremal black hole geometry. By varying this relation, it is then easy to verify the first law of near horizon mechanics
\be 
\frac{\dd A}{8\pi}=\mu \dd Q\,,
\ee 
where $\mu=Q$ is the conjugate potential, obtained by the regularity of the gauge field~\cite{Hajian:2013lna, Hajian:2014twa}. It is easy to verify that the same law can also be obtained from the full static (extremal) geometry.

\section{Extremal rotating black holes in Einstein--Maxwell}\label{Sec:4}

To set the stage for the analysis of Einstein--Born--Infeld theory to be considered in the next section, we begin with a study of the extremal rotating black holes in Einstein--Maxwell theory. In this case, the full bulk solution is known analytically (the Kerr--Newman solution) and the equations of the near horizon geometry can be solved analytically. Hence, this serves as a useful check of our formalism and aids in understanding the methodology.

\subsection{Extremal Kerr--Newman}

The Kerr--Newman metric  \cite{Newman:1965my, Newman:1965tw} is an exact solution of the Einstein--Maxwell equations. Here we briefly review the extremal limit of this metric and the relationship between the physical parameters at extremality.

The Kerr--Newman metric and gauge field take the following form:
\begin{align}
{\rm d}s^{2} ={}&
- \frac{\Delta}{\rho^{2}}\!\left({\rm d}t - a\sin^{2}\theta\,{\rm d}\phi\right)^{2}
+ \frac{\sin^{2}\theta}{\rho^{2}}
      \left[(r^{2}+a^{2})\,{\rm d}\phi - a\,{\rm d}t\right]^{2} + \frac{\rho^{2}}{\Delta}\,{\rm d}r^{2}
+ \rho^{2}\,{\rm d}\theta^{2}\,,\nonumber
\\
A
={}& -\frac{Q\,r}{\rho^{2}}\left({\rm d}t - a\sin^{2}\theta\,{\rm d}\phi\right),
\end{align}
where
\begin{align}
\rho^{2} &= r^{2} + a^{2}\cos^{2}\theta,  \qquad 
\Delta   = r^{2} - 2 M r + a^{2} + Q^{2}.
\end{align}
Here $a = J/M$ is the angular momentum per unit mass while $Q$ is the electric charge.

The horizon structure of the Kerr--Newman metric is determined by the roots of $\Delta$. It is ``Reissner--Nordstr{\"o}m-like'' -- we find up to two horizons:
\be 
r_\pm=M\pm \sqrt{M^2-a^2-Q^2}\,.
\ee 
The black hole is extremal, 
when the two roots coincide (in which case the black hole temperature vanishes, $T=0$). It  happens when
\be 
r_+=r_-=M=\sqrt{a^2+Q^2}\,.
\ee
We can express the result in a form more suitable for comparison if we write $J = J(Q, A)$ where $A$ is the horizon area. Since $A = 4 \pi (r_+^2 + J^2/M^2)$ we then have
\be 
A^2=(8 \pi J)^2 + \left(4 \pi Q^2\right)^2 \, .
\ee
Thus, it is easy to see that, at fixed area, the phase space of extremal solutions consists of a circle in the $(Q, J)$ plane. Below, we will see how to reproduce this relationship using only the near horizon limit of the extremal solution.

\subsection{Extremal near horizon metric approach}

Let us now  illustrate how the equations of motion are solved for the extremal near horizon geometry given in eq.~\eqref{eq:ext_geom}. Taking $\mathcal{L}_{\rm Max} = -\mathcal{S}/2$, the Einstein--Maxwell equations~\eqref{eq:ein_nle} admit the following exact solution:
\begin{align}\label{MaxwellSolution}
    h(y) &= - \frac{h_0 \left(y^2 x_0^2 - n^2 \right)}{y^2 x_0^2 + n^2 } + \frac{n^2 h_1 y}{y^2 x_0^2 + n^2} \,,\nonumber
    \\
    g(y) &= \frac{y(l-y) x_0^4 + n^2x_0^2(1-4h_0^2)-h_1^2 n^4}{x_0^4 \left(y^2 x_0^2 + n^2 \right) } \, ,
\end{align}
where $h_0=h(0), h_1=\frac{\dd h}{\dd y}(0)$, and $l=n^2 g_1=n^2g'(0)$ are constants of integration.
The constant $l$ is actually a NUT charge, and it breaks the $y \to - y$ symmetry of $g(y)$. Hence, we will set this parameter to zero. To interpret $h_0$ and $h_1$ we evaluate the electric and magnetic charges:
\begin{align}
Q_e &= \frac{1}{4 \pi} \int \star D = - \frac{2n x_0 h_0}{\omega \left( x_0^2 + n^2 \right)} \, ,
\\
Q_m &= \frac{1}{4 \pi} \int F = \frac{n^2 h_1}{\omega \left(n^2 + x_0^2\right)} \, .
\end{align}
We therefore see that $h_0$ is related to the electric charge while $h_1$ is related to the magnetic charge. For the sake of simplicity, we will take $h_1=0$, setting $Q_m = 0$ henceforth (and shall refer to $Q_e$ as simply $Q$). On the other hand, we can compute the angular momentum using the modified Komar integral defined above. Using the regular gauge field, we find
\be 
J = \frac{(1-4h_0^2) x_0 n^3 + x_0^3 n}{2 \omega^2 \left( n^2 + x_0^2 \right)^2} \, .
\ee

Note that the constants in the solution are not all free but are constrained by regularity \eqref{eq:regularity}. In particular, we must demand that $g(\pm 1) = 0$. With $l= 0$, this is actually just a single constraint since $g(-y) = g(y)$. Hence, solving for $n$ and $\omega$ as a function of the other parameters we find,
\be 
n = \sqrt{Q^2 + x_0^2} \, , \quad \omega = \frac{x_0}{Q^2 + 2 x_0^2} \, .
\ee
Plugging these into the expression for the area we obtain,
\be 
A = 4 \pi \left(Q^2 + 2 x_0^2 \right) \, .
\ee
We can solve this expression for $x_0$ and then plug all of the above into the expression for the angular momentum to obtain an explicit relationship between the physical properties of the solution:
\be \label{eq:JQA_max}
(8 \pi J)^2 + \left(4 \pi Q^2 \right)^2 = A^2 \, .
\ee
The result is exactly the same as we obtain working from the full Kerr--Newman solution (see previous subsection). Thus, we have obtained a valuable consistency check of our methods. Note that the extremal solutions completely fill the $(Q, J)$ plane. The expression is symmetric under $Q \to -Q$ and $J \to -J$; for a fixed value of the area, the constraint traces out a circle.

One of our goals in the analysis that follows is to understand how the space of extremal rotating solutions changes when Born--Infeld electrodynamics is considered. We already know that the $J = 0$ slice of this space is drastically modified, with extremal solutions existing only for $Q \ge 1/(2b)$. 
The question is whether any rotating extremal solutions are also excluded. 

\section{Extremal rotating black holes in Einstein--Born--Infeld}
\label{Sec:5}

\subsection{Equations of motion and exact solution}\label{sec:eqns_and_sol}

\iftrue
To simplify the form of the following expressions, here we define two quantities that absorb various parameter dependencies.
\be \label{eq:tu_def}
t \equiv \frac{x_0 y}{n} \, , \quad u \equiv \left(\frac{h_0}{n b}\right)^2\,.
\ee
With the ansatz introduced earlier for the gauge field, there is a single nontrivial term for the Born--Infeld equations:
\be 
 \Gamma h''+2 t \left(t^2+1\right) h'+4 h= 0 \, ,
\ee
where $h=h(t)$, $h'=\dv{h}{t}$, and we introduced the notation
\be 
\Gamma = \frac{b^2 n^2 \left(t^2+1\right)^2-4 h^2}{b^2 n^2+h'^2} \, .
\ee
There is also a single nontrivial term for the Einstein equations. All other components can be reduced to this equation (in some cases requiring also the use of the electromagnetic equations). We can write this in the following way: 
\be \label{bi_efe}
 t g'+g\frac{t^2-1}{t^2+1} +1= \frac{8 \pi  n^2}{ x_0^2} \left(t^2+1\right) g T_{yy}\,,
\ee
where $g=g(t)$, $g'=\dv{g}{t}$, and $T_{yy}$ is the corresponding component of the Born--Infeld stress tensor: 
\be 
T_{yy} = \frac{b^2 x_0^2}{4 \pi g}\left(1-\frac{\sqrt{\Gamma}}{t^2+1}\right)\,.
\ee

Remarkably, we were able to find an {\em exact solution} to the above equations. As before, we have denoted $h_0=h(0)$ and set $ h'(0)=0=g'(0)$. The solution then reads\footnote{
One helpful fact in obtaining the solution is to note that the \textit{local form} of the near horizon metric of the rotating black hole~\eqref{eq:ext_geom} is equivalent, after a double Wick rotation, to the hyperbolic Taub-NUT geometry. While we are not aware of Taub-NUT metrics constructed in Born--Infeld theory for general base spaces, the case of a spherical base appears in~\cite{GarciaD:1984xrg,Breton:2014mba}. While this is not what we need here, the structure of the solution is qualitatively similar. 
Due to the analogy with the Taub-NUT metrics, we expect that it may be possible to construct analogous solutions for some other theories of NLE, e.g. \cite{BallonBordo:2020jtw}. 
Note, however, that even though the \textit{local form} of the metrics can be mapped via double Wick rotation into hyperbolic Taub-NUT solutions, the \textit{global properties} of the solution (such as their regularity and conserved charges) will be wholly different -- analogous to the differences between the AdS black brane and the AdS soliton which are similarly related via double Wick rotation.}
\begin{align}\label{eq:exact_sol}
    h(y) &= h_0 \cosh \left(\frac{2 F(\Phi |m)}{\beta}\right) \, ,\nonumber
   \\
   g(y) &= \frac{1-t^2}{x_0^2 \left(1+t^2\right)} + \frac{2 h_0^2}{3 u x_0^2 \Delta \left(1+t^2\right)  } \bigg[3+5 t^2+t^4-t^6+4 u\left(-3+t^2\right)+\Delta  \left(-3+6 t^2+t^4\right)
   \nonumber
   \\
   &\qquad +\frac{8 i t \Delta  \left(\beta ^2 E(\Phi |m)+\left(2-\sqrt{u}\right)
   \sqrt{u} F(\Phi |m)\right)}{\beta }\bigg]\,,
\end{align}
where $F(\Phi|m)$ is the elliptic integral of the first kind, $E(\Phi|m)$ is the elliptic integral of the second kind,\footnote{Our conventions for the elliptic integrals match those of Mathematica, namely: $F(\Phi|m) = \int_0^\Phi \left(1-m \sin^2\theta\right)^{-1/2}{\rm d}\theta$ and $E(\Phi|m)=\int_0^\Phi \left(1-m \sin^2\theta\right)^{1/2}{\rm d}\theta$.} and we introduced
\be\label{eq:defns} 
\Delta = \sqrt{\left(t^2+1\right)^2-4 u} \, , \quad \Phi = i \sinh ^{-1}\left(\frac{t}{\sqrt{1+2 \sqrt{u}}}\right) \, , \quad m = \frac{1+2 \sqrt{u}}{1-2 \sqrt{u}} \, , \quad  \beta = \sqrt{1-2 \sqrt{u}} \, .
\ee
Similar to the Einstein--Maxwell solution \eqref{MaxwellSolution} (after eliminating $g_1$ and $h_1$), our solution depends on three parameters $(h_0, x_0, n)$.\footnote{The parameter $\omega=\mp x_0g'(\pm 1)/2$ can be expressed as a function of the above three parameters, $\omega=\omega(h_0, x_0, n)$ by employing the equation of motion \eqref{bi_efe}.}
However, not all of these parameters can be independent. To ensure the regularity conditions~\eqref{eq:regularity} are satisfied, one of these parameters must be a function of the remaining two. It is not possible to obtain this relationship analytically, and so we shall use a combination of perturbative and numerical methods to extract meaningful physical information from these equations.

Before concluding this subsection, let us write the integral form of the relevant charges. These read:
\begin{align}
    Q&\equiv Q_e=  -\frac{x_0}{n\omega} \int_{-\frac{x_0}{n}}^{\frac{x_0}{n}} \frac{h(t)}{\sqrt{\Gamma}} \, {\rm d} t\, ,\qquad  
   Q_m = \frac{h(\frac{x_0}{n}) - h(-\frac{x_0}{n})}{2 \omega}=0 \, ,\nonumber\\
    J & = -\frac{x_0}{4  n \omega ^2}\int_{-\frac{x_0}{n}}^{\frac{x_0}{n}} \left(\frac{4 h(t) (h(t)-h(\frac{x_0}{n}))}{\sqrt{\Gamma}}+\frac{g(t)}{t^2+1}\right)  \, {\rm d} t\, .
\end{align}
We shall use these formulae in the sections to follow.

\subsection{Perturbative results}

The general solution presented above is rather opaque due to the nature of the special functions appearing in it. Moreover, despite having a closed form expression in terms of the parameters $(h_0, x_0, n)$, the integrals defining the physical charges cannot be evaluated. Hence, it will be useful to expand the solution in different limits to extract results with more physical clarity. We begin by first expanding the solution in the limit of small angular momentum.

\subsubsection{Expansion in small angular momentum} \label{sec:small_J}

As we saw in the example with Einstein--Maxwell theory, the parameter $x_0$ is related to the spin parameter of the Kerr--Newman solution. Hence, we begin our perturbation journey by constructing solutions in the limit of small $x_0$. In this limit, the functions $h(y), g(y)$ admit the following expansions:
\be 
g(y) = \sum_i x_0^i \, g^{(i)}(y)  \,  ,\quad h(y) = \sum_i x_0^i \,  h^{(i)}(y)  \, ,
\ee
For the function $h(y)$, the first few expansion coefficients are\footnote{Here, we have already eliminated $h_0$ by demanding the regularity condition $g(\pm 1)=0$.} 
\begin{align}
    h^{(0)} =&\, -\frac{\sqrt{4 b^2 n^2-1}}{4 b n} \, ,\nonumber
    \\
    h^{(2)} =&\, \frac{b}{2 n \sqrt{4 b^2 n^2-1}} \left(1-\frac{4 b^2 n^2 y^2 \left(1-4 b^2 n^2\right)}{\left(1-2 b^2 n^2\right)^2}\right) \, ,
\end{align}
while for the function $g(y)$ we have
\begin{align}
    g^{(0)} =& \, \frac{2 b^2 \left(1-y^2\right)}{2 b^2 n^2-1} \, ,\nonumber
    \\
    g^{(2)} =& \, \frac{2 b^2 \left(y^2-1\right) \left(12 b^4 n^4 y^2+3 b^2 n^2 \left(5-3 y^2\right)+2 \left(y^2-2\right)\right)}{3 n^2 \left(2 b^2 n^2-1\right)^3} \, .
\end{align}

We have computed the terms in this expansion to much higher order, but the resulting expressions are somewhat unwieldy.
With the terms in the expansion at hand, it becomes a straightforward exercise to evaluate the charge and angular momentum. In terms of $(x_0, n)$ the results are
\begin{align}
    Q &= \frac{\sqrt{4 b^2 n^2-1}}{2 b}+\frac{\left(-24 b^6 n^6+24 b^4 n^4-14 b^2 n^2+2\right) x_0^2}{6 b \sqrt{4 b^2 n^2-1} \left(n-2 b^2 n^3\right)^2} 
    \nonumber
    \\
    &+\frac{\left(-593 b^4 n^4+94 b^2 n^2-24 b^6 n^6 \left(30 b^6 n^6-180 b^4 n^4+119 b^2 n^2-68\right)-5\right) x_0^4}{45 b \left(4 b^2
   n^2-1\right)^{3/2} \left(n-2 b^2 n^3\right)^4} + \mathcal{O}(x_0^6) \, ,
   \\
   J &= \frac{ \left(12 b^4 n^4-6 b^2 n^2+1\right) x_0}{6 b^2 n \left(2 b^2 n^2-1\right)} -\frac{\left(144 b^6 n^6+168 b^4 n^4-66 b^2 n^2+5\right)x_0^3}{90 b^2 n^3 \left(2 b^2 n^2-1\right)^3} + \mathcal{O}(x_0^5) \, .
\end{align}

A crucial fact about the above series solution is that its validity requires 
\be\label{n2} 
n^2 > 1/(2 b^2)\,.
\ee 
As we shall see later, this bound arises only in the small $x_0$ expansion and is not a constraint on the full solution presented in the previous section.  However, it is now easy to see that it generalizes the bound $|Q| > 1/(2b)$, we had for static solutions -- recovered in the limit $x_0\to 0$.  

We can now go further and invert the above expressions to obtain $(n, x_0)$ as functions of $(Q, J)$. For convenience, we record the result of this inversion here:
\begin{align}
    n =&\, \frac{\sqrt{4 b^2 Q^2+1}}{2 b}+\frac{12 b^3 J^2 \left(192 b^6 Q^6-48 b^4 Q^4+52 b^2 Q^2+3\right)}{\sqrt{4 b^2 Q^2+1} \left(48 b^4 Q^4+1\right)^2} + \mathcal{O}(J^4) \, ,
    \\
    x_0 =&\, \frac{6 b J \left(4 b^2 Q^2-1\right) \sqrt{4 b^2 Q^2+1}}{48 b^4 Q^4+1}
    \nonumber
    \\
    &-\frac{144 b^5 J^3 \left(8 b^2 Q^2 \left(-986 b^2 Q^2+96 b^4 Q^4 \left(240 b^6 Q^6-216 b^4 Q^4-179 b^2
   Q^2-19\right)+35\right)+25\right)}{5 \sqrt{4 b^2 Q^2+1} \left(48 b^4 Q^4+1\right)^4} 
   \nonumber
   \\
   &+ \mathcal{O}(J^5) \, .
\end{align}
Then we are able to express the area (and hence semiclassical entropy) of the extremal horizon as a function of the physical charges. Despite the complexity of some of the intermediate formulae, the leading behaviour of the area is rather simple, reading
\begin{align}
    \label{horizon_area}
    A &= \frac{\pi  \left(4 b^2 Q^2-1\right)}{b^2}+\frac{96 \pi  b^2 \left(4 b^2 Q^2+1\right) J^2}{1+48 b^4 Q^4} + \mathcal{O}(J^4) \, .
\end{align}
Higher order terms rapidly become complicated --- we present a few more of these in the appendix. We emphasize that while this result is perturbative in the angular momentum, it is \textit{exact} in the charge and Born--Infeld parameter. 

Besides studying the relationship between the area and the physical charges $(Q, J)$, there is another use for this expression we have obtained. Recall in the case of the spherically symmetric black holes that extremal solutions exist only for $|Q| > 1/(2b)$. Precisely when this bound is saturated, the area of the extremal horizon vanishes. We can apply similar reasoning to the rotating extremal black holes. By solving for $Q$ as a function of $J$ such that the area \textit{vanishes} we are likely to obtain a useful boundary in the parameter space of the extremal rotating solutions. Carrying out this calculation, we find that
\begin{align}
    \label{area_charge}
    b |Q_\star^{(J)}| =& \, \frac{1}{2}-12 j^2-\frac{3024 j^4}{5}-\frac{8843904 j^6}{175}-\frac{4476135168
   j^8}{875} + \mathcal{O}(j^{10}) \, ,
\end{align}
where $j \equiv J b^2$ is a dimensionless angular momentum. We include higher powers in the expansion in the appendix. As is clear from an examination of the coefficients, they grow very rapidly\footnote{The growth is faster than $n!$ but slower than $(2n)!$. Hence, if a closed form expression could be obtained, a second-order Borel summation may be helpful.} and the series has vanishing radius of convergence. Nonetheless, we find that certain resummation techniques exhibit rapid convergence. In particular, we have found that diagonal Pad{\'e} approximants are especially useful. For example, the $[2|2]$ Pad{\'e} approximant
\be 
b |Q_\star^{(J)}| \approx \frac{5-372 j^2}{10-504 j^2}
\ee
has a maximum absolute error of $0.05$ for $|j| < 0.081$ compared to the highest order Pad{\'e} approximants we have constructed. It therefore gives a simple and reasonable estimate of the value of charge for which the area of the extremal horizon vanishes. 

The conclusion of the above analysis is that the perturbative results strongly suggest that there is a curve $Q_\star^{(J)}(J)$, extending the (zero angular momentum) `star point' $Q_\star=Q_\star^{(J)}(0)$, in the solution space for which the area of the horizon vanishes. When scanning 
the parameter space in more detail numerically, we shall see that this feature is indeed borne out.

\subsubsection{Expansion in Born--Infeld parameter} \label{sec:large-b}

To better understand how the Born--Infeld corrections affect the Maxwellian results, we expand the closed form solution for large $b$. In the limit $b \to \infty$, the functions admit the following expansions: 
\be 
h(y) = - \frac{h_0 \left(y^2 x_0^2 - n^2 \right)}{y^2 x_0^2 + n^2} +  \sum_{k = 1}^\infty \frac{h^{(2k)}(y)}{b^{2k}} \, , \quad g(y) = \frac{(1-4h_0^2)n^2 - y^2 x_0^2}{x_0^2 \left(y^2 x_0^2 + n^2 \right)} + \sum_{k = 1}^\infty \frac{g^{(2k)}(y)}{b^{2k}} \, .
\ee
We then plug these expansions into the equations of motion, expand in the limit of large $b$, and solve order-by-order. 

Let us consider first the solution for $h(y)$. The first few terms in this expansion read,
\begin{align}
    h^{(2)}(y) &= -\frac{t^2 h_0^3  \left(3 t^2+5\right)}{n^2 \left(t^2+1\right)^3}-\frac{3 t h_0^3}{n^2
   \left(t^2+1\right)} \arctan(t)\, ,\nonumber
    \\
    h^{(4)}(y) &= -\frac{t^2 \left(329+521 t^2+343 t^4+87 t^6\right) h_0^5}{16 n^4 \left(1+t^2\right)^5}-\frac{3 t \left(55+62
   t^2+23 t^4\right) h_0^5}{16 n^4 \left(1+t^2\right)^3}\arctan(t)
   \nonumber
   \\
   &+\frac{9 (-1+t) (1+t) 
   h_0^5}{8 n^4 \left(1+t^2\right)} \arctan^2(t) \, ,
\end{align}
where we are expressing the results in terms of the parameter $t$ introduced in eq.~\eqref{eq:tu_def} which dramatically simplifies the expressions. It is straightforward to continue the computation to much higher order but the results rapidly become very complicated. We note that the general structure of the $h^{(2k)}(y)$ terms is clear and they have the schematic form
\be 
h^{(2k)}(y) = \frac{h_0^{2k+1}}{n^{2k}} \sum_{i=0}^{k} \frac{R_{i, k}(t) }{\left(1+t^2\right)^{2k+1-2i}} \arctan^k\left(t\right) \, , 
\ee
where $R_{i, k}(t)$ are a particular set of polynomials in the parameter $t$.\footnote{For example, $R_{0, 1} = -t^2(3t^2+5)$ and $R_{1,1} = -3 t$. } However, despite this clear structure it is hard to understand the general pattern for these terms in the expansion. 

On the other hand, the expansion of $g(y)$ has a simple structure at all orders in the large $b$ expansion, and we can express the general result in different ways. Expressing it in integral form we have,
\be 
g^{(2k)}(y) = \frac{4 h_0^2 t}{x_0^2 \left(1+t^2\right)}  \left(\frac{h_0}{n} \right)^{2k}\frac{(2k)!}{k! (k+1)!} \int \frac{{\rm d}t}{t^2 \left(1 + t^2 \right)^{2k}} \, ,
\ee
or in terms of hypergeometric functions we have
\be 
g^{(2k)}(y) = - \frac{4 h_0^2 }{ x_0^2 \left(1+t^2\right)} \left(\frac{h_0}{n} \right)^{2k} \frac{(2k)!}{k! (k+1)!}  \, _2F_1\left(-\frac{1}{2},2
   k;\frac{1}{2};-t^2\right) \, .
\ee

With these perturbative expressions in hand, we can evaluate the physical properties of the black hole. Defining
\be 
\alpha \equiv \frac{A}{4 \pi Q^2} \, ,
\ee
we find that the relationship between the angular momentum, horizon area, and charge takes the form
\begin{align}
    \frac{J}{Q^2} =& \, \frac{ \sqrt{\alpha^2 - 1}}{2} + \frac{(4 \alpha +1) \sqrt{\alpha ^2-1}+6 \alpha ^2 \arctan \left(\sqrt{\frac{\alpha-1}{\alpha +1}}\right)}{16 (bQ)^2 (\alpha -1) (\alpha +1)^3} 
    \nonumber
    \\
    &+  \frac{1}{(bQ)^4 (\alpha -1)^2 (\alpha +1)^6 \sqrt{\alpha ^2-1} } \bigg[-\frac{9}{16} \alpha ^2 (\alpha  (3 \alpha -4)+2) \arctan^2\left(\sqrt{\frac{\alpha -1}{\alpha
   +1}}\right)
   \nonumber
   \\
   &+\frac{1}{64} \alpha  \sqrt{\alpha ^2-1} ((\alpha -1) \alpha  (35 \alpha
   -121)-12) \arctan\left(\sqrt{\frac{\alpha -1}{\alpha +1}}\right)
   \nonumber
   \\
   &+\frac{(\alpha -1) (\alpha
   +1) (\alpha  (\alpha  (\alpha  (160 \alpha -261)+76)+15)+4)}{384 \alpha } \bigg] + \mathcal{O}(b^{-6}) \, ,
\end{align}
which generalizes eq.~\eqref{eq:JQA_max} to include perturbative Born--Infeld corrections. We note that for all $\alpha > 1$, the leading correction is strictly positive, while the next-to-leading correction is strictly negative. The higher-order terms in the series continue to alternate in the same manner. The similarity between the expansion for $J$ and the expansion for $h(y)$ in this limit (both consisting of rational functions plus polynomials in $\arctan$) make it tempting to speculate that a closed form expression could exist for the angular momentum. However, we have not been able to find such an expression.

\subsection{First law of near horizon mechanics}

Throughout this work, we have constructed and studied the physical properties of the near horizon geometries of rotating Einstein--Born--Infeld black holes. While we do not have access to the asymptotic region of these solutions, it is nonetheless possible to study an analog of the first law of black hole mechanics suitably adapted to the near horizon regime. Specifically, the first law for near horizon geometries reads~\cite{Hajian:2013lna, Hajian:2014twa, Cassani:2023vsa, Cano:2024tcr}
\be 
\frac{{\rm d}A}{8 \pi} = \varpi \, {\rm d J} + \mu \, {\rm d Q} \, .
\ee
Here $A, J, Q$ are the horizon area, angular momentum, and charge. It must be emphasized that the conjugate potentials $\varpi$ and $\mu$ do not bear a simple relationship to the angular velocity or electric potential that would enter into the first law that applies to the full, asymptotically flat spacetime. As explained in~\cite{Hajian:2013lna, Hajian:2014twa}, the potentials $\varpi$ and $\mu$ can be obtained as singular limits of the angular velocity and electric potential, but it is difficult to see how this limit could be `undone' (indeed, it is akin to understanding how the near horizon geometry is contained in a full asymptotically flat solution). 

The conjugate potentials $\varpi$ and $\mu$ are fixed from the near horizon metric and gauge field by demanding its regularity -- see~\cite{Cano:2024tcr} for a detailed discussion of this in the context of higher-dimensional rotating black holes.  In terms of the quantities appearing in our near horizon metric and gauge field we have,
\be 
\varpi = 2 n \omega \, , \quad \mu = -2 n h(1) \, .
\ee
Since the charges of our full solution, complete with the regularity condition, can only be handled numerically, we shall verify the first law perturbatively, working in the limit of large Born--Infeld parameter. In this limit, the relevant physical quantities are to leading order as follows:
\begin{align}
    J &=  \frac{\sqrt{A^2 - (4 \pi Q^2)^2}}{8 \pi} +   \frac{4 \pi ^2 Q^4 \left(A+\pi  Q^2\right) \sqrt{A^2-16 \pi ^2 Q^4}}{b^2 \left(A-4 \pi  Q^2\right) \left(A+4
   \pi  Q^2\right)^3}
   \nonumber 
   \\
   &+\frac{6 A^2 \pi ^2 Q^4 }{b^2
   \left(A-4 \pi  Q^2\right) \left(A+4 \pi  Q^2\right)^3} \arctan \left(\sqrt{-1+\frac{2 A}{A+4 \pi  Q^2}}\right) + \mathcal{O}(b^{-4}) \, ,
    \\
    \varpi &= \frac{\sqrt{A^2 - (4 \pi Q^2)^2}}{A} + \frac{64 \pi ^3 Q^4 \left(A-2 \pi  Q^2\right)^2}{b^2 A^2 \sqrt{A-4 \pi  Q^2} \left(A+4 \pi 
   Q^2\right)^{5/2}}
   \nonumber
   \\
   &+\frac{96 \pi ^3 Q^4 \left(A^2-4 A \pi  Q^2+16 \pi ^2 Q^4\right) }{b^2 A  \left(A-4 \pi  Q^2\right) \left(A+4 \pi 
   Q^2\right)^3} \arcsin \left(\sqrt{\frac{1}{2}-\frac{2 \pi  Q^2}{A}}\right) + \mathcal{O}(b^{-4}) \, ,
    \\
    \mu &= \frac{4 \pi Q^3}{A} -\frac{16 \pi ^2 Q^3 \left(A-2 \pi  Q^2\right)^2}{b^2 A^2  \left(A+4 \pi  Q^2\right)^2}
    \nonumber
    \\
    &-\frac{24 \pi ^2 Q^3 \sqrt{A^2-16 \pi ^2 Q^4} \left(A^3+64
   \pi ^3 Q^6\right) }{b^2 A \left(A-4 \pi  Q^2\right) \left(A+4 \pi 
   Q^2\right)^4}\arcsin \left(\sqrt{\frac{1}{2}-\frac{2 \pi  Q^2}{A}}\right)+ \mathcal{O}(b^{-4}) \, .
\end{align}
One can then verify the first law order-by-order in a large $b$ expansion. The zeroth order contributions coincide with those of the Kerr--Newman near horizon geometry, which of course satisfies the near horizon first law. With the expressions given above, it is straightforward to verify that the first law holds up to $\mathcal{O}(b^{-4})$. We have carried out the procedure to higher order in Mathematica, finding that the first law is satisfied in all cases we have assessed. 

It must be noted that the first law is not assumed in the derivation of any quantity which enters it. Therefore, the fact that it holds provides a highly nontrivial consistency check of all our expressions and results.

\subsection{Numerical exploration of solution space}

To obtain a full picture of the solutions and their properties we will need to resort to numerical methods, working from the analytical solution \eqref{eq:exact_sol}. Numerical techniques are required for two aspects of the problem: (1) enforcing the regularity condition which relates one parameter from the set $\{h_0, x_0, n\}$ to the other two, and (2) performing the numerical integration required to obtain the angular momentum and electric charge. Both of these problems are straightforward and we can use the built in functions of Mathematica (such as \textsc{FindRoot} and \textsc{NIntegrate}) for these purposes. At the same time, we have also performed a numerical consistency check of the closed form solution we obtained in eq.~\eqref{eq:exact_sol} by the full numerical integration of the equations of motion -- this appears in appendix~\ref{app:numerical_check}.


\subsubsection{General remarks}

Before diving into the numerical exploration of the solution space, let us summarize a few points that can be learned from the equations, the solutions, and the perturbative analyses we conducted earlier. 
\begin{itemize}
    \item The equations of motion are invariant under $x_0 \to - x_0$ and $n \to -n$ and also under $h(y) \to - h(y)$. Hence, we can restrict ourselves to one sign of the parameters $(h_0, x_0, n)$ if desired. Note, however, that the charges are \textit{not} invariant under these transformations. For example, $h(y) \to - h(y)$ changes the sign of the electric charge while $n \to -n$ changes the sign of the angular momentum. Changing $x_0 \to - x_0$ simply switches the north and south poles of the 
    sphere. Hence, we shall take $x_0 > 0$ without loss of generality. 
    \item From the small angular momentum expansion discussed in section~\ref{sec:small_J}, we saw that constraint $n^2 \ge  1/(2b^2)$ played an important role. In the perturbative regime, this bound, in some sense, generalized the $Q > 1/(2b)$ constraint required for the existence of extremal static black holes. Here we shall see that the curve 
    \be 
    n = 1/(\sqrt{2} b)
    \ee 
    plays an important role in the general situation, marking the onset of {\em strong nonlinear effects}.
    
    \item From the exact solution presented in section~\ref{sec:eqns_and_sol}, it is obvious, see Eq. \eqref{eq:defns}, that the 
    existence of the solution requires that $\beta>0$, i.e., 
    \be 
    h_0^2 < \Bigl(\frac{b n}{2}\Bigr)^2\,.
    \ee 
    
     We will find that when $n > 1/(\sqrt{2} b)$ the inequality is \textit{strict}, while we find evidence that the inequality is saturated (allowing the parameter $m\to \infty)$ as $x_0 \to 0$ for $n < 1/(\sqrt{2} b)$.
\end{itemize}

\begin{figure}
    \centering
    \includegraphics[width=0.75\linewidth]{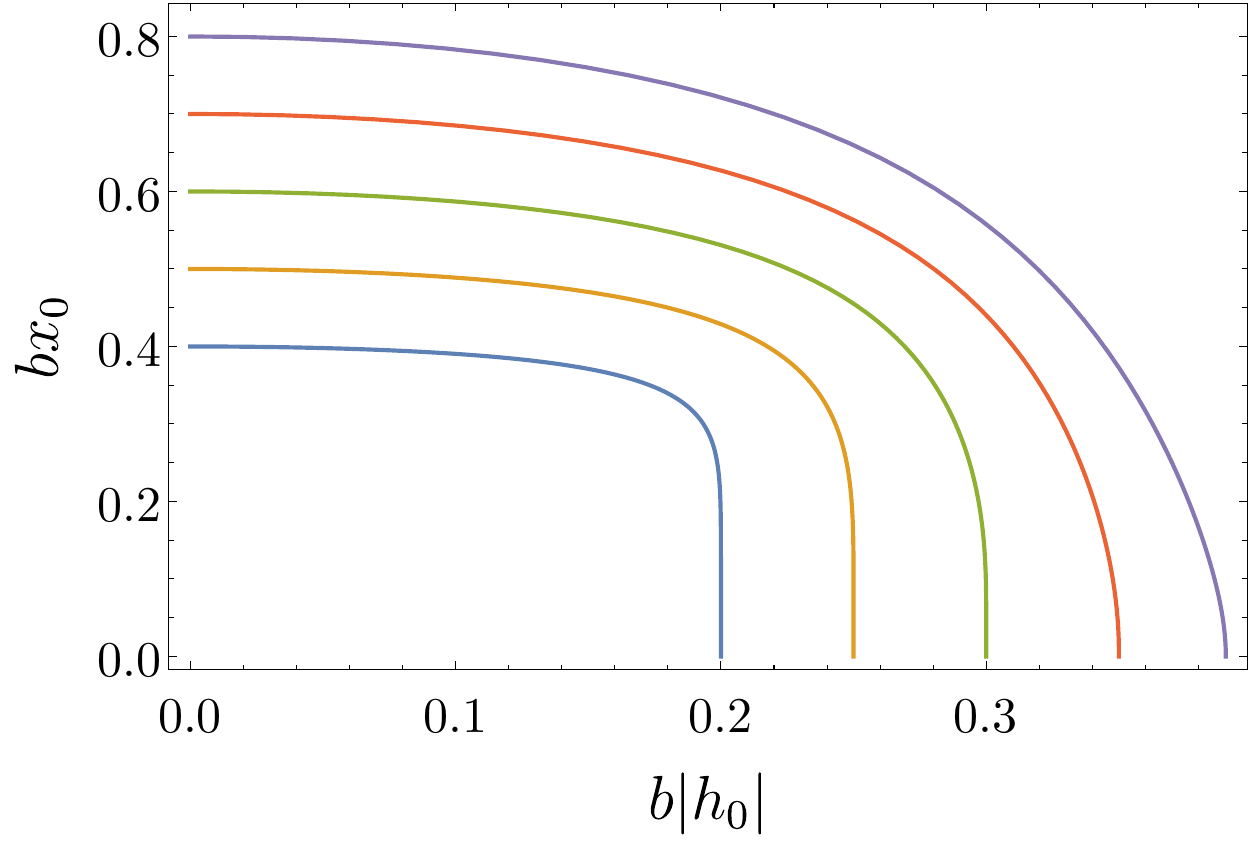}
    \caption{{\bf Parameter constraints from regularity.} The parameter $x_0$ is plotted as a function of $|h_0|$. The different curves correspond to $n = 0.4, 0.5, 0.6, 0.7, 0.8$ in order of lower curves to higher ones; the curves intersect the vertical axis at $x_0 = n$. Here we have set the units such that $b=1$.}
    \label{fig:x0_curves}
\end{figure}

The general approach we shall take here is to regard $(h_0, n)$ as specified parameters and $x_0$ as a \textit{derived} parameter obtained by enforcing the regularity condition $g(\pm 1) = 0$. In principle, one could imagine that for fixed $(h_0, n)$ there may be multiple solutions $x_0$. However, this is not what we observe -- we find either a \textit{unique} solution $x_0(h_0, n)$ or \textit{no} solution. 

We illustrate in Figure~\ref{fig:x0_curves} the dependence of $x_0$ on $h_0$ for different choices of $n$, here taken to be $n = 0.4, 0.5, 0.6, 0.7, 0.8$ in units such that $b = 1$.\footnote{Throughout this section, anytime a dimensionful quantity is referred to it is to be understood we are working in units such that $b = 1$.} We have in all cases $x_0 \to n$ as $h_0 \to 0$. This fact is provable analytically -- it follows simply from the fact that the electric charge vanishes in this limit and the resulting solution is just near horizon extremal Kerr. As $|h_0|$ increases from zero, all  curves exhibit the same qualitative structure, with $x_0$ \textit{decreasing} toward zero from its initial value of $n$. However, the rate at which the curves limit to zero strongly depends on the value of $n$. When $n$ is large, the curves gradually curve toward $x_0 = 0$ as $|h_0|$ increases. When $n$ is small [specifically, when $n < 1/(\sqrt{2} b)$], $x_0$ initially exhibits a very weak dependence on $h_0$, but then suddenly drops rapidly toward $x_0 = 0$ as $|h_0| \to bn/2$. In the figure, this is most pronounced for the $n = 0.4$ curve. 

From a practical point of view, this means that for small values of $n$ it becomes increasingly difficult to push the numerical solutions to very small $x_0$ as it requires very high precision. For example, let us consider the case $n = 0.4$, for which the absolute upper limit of $|h_0|$ is $0.2$. We find that when $|h_0| = 0.19$ we have $x_0 \approx 0.3144891675$, when $|h_0| = 0.2 -10^{-7}$ we have $x_0 \approx 0.1030305771$, and when $|h_0| = 0.2 - 10^{-20}$ we have $x_0 \approx 0.03094903365$. In fact, to have $x_0 < 10^{-3}$ requires $|h_0| \approx 0.2 - 10^{-500}$! This illustrates just how sensitive the dependence becomes, and it becomes even more sensitive the smaller $n$ is taken. A bit further below, we shall return to this issue from an analytic viewpoint.

\begin{figure}
    \centering
    \includegraphics[width=\linewidth]{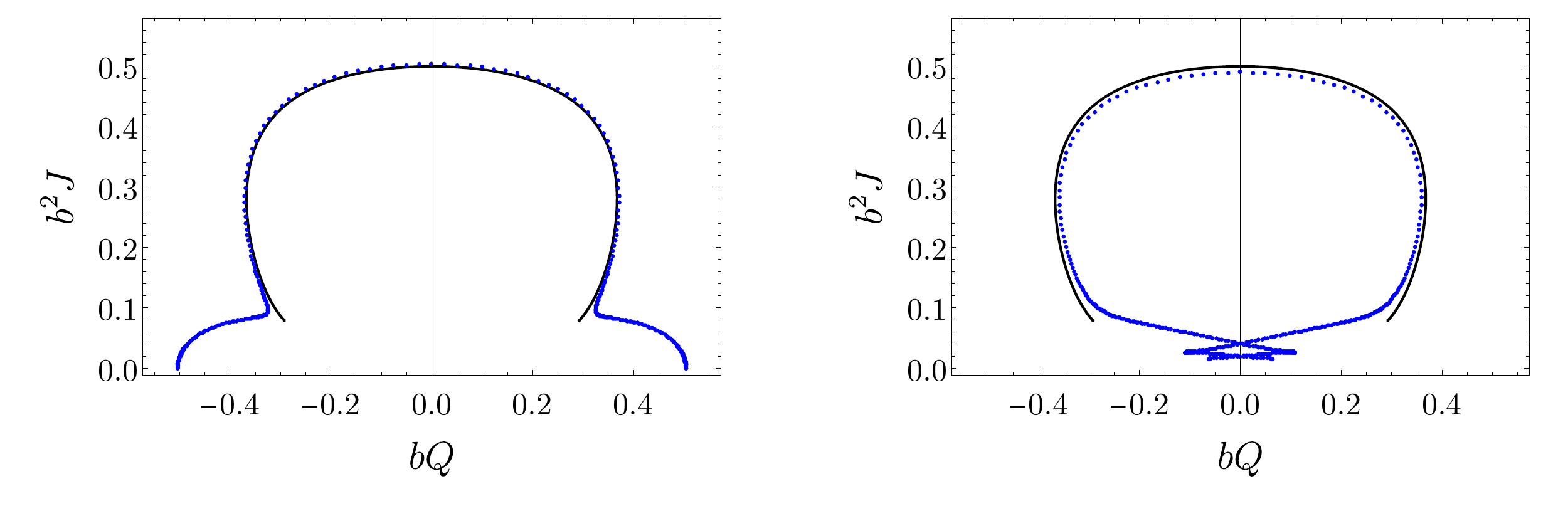}
    \caption{{\bf Two distinct behaviors near $n=1/\sqrt{2}$.} We show the two types of behaviours in the $(Q, J)$ parameter space for extremal solutions. In each plot, the black curve corresponds to $n = 1/\sqrt{2}$ while each blue dot corresponds to solutions with different values of $h_0$. The left plot shows the case of 
    $n=0.71$
    while the right plot shows 
    $n=0.7$. 
    Units are such that $b = 1$.}
    \label{fig:example_curves}
\end{figure}

\subsubsection{
Parameter space and physical implications}

Let us now explain how the solutions behave in the $(Q, J)$ parameter space. 
We illustrate this in Figure~\ref{fig:example_curves}, which shows the curve $n = 1/\sqrt{2}\approx 0.70711$ (the black curve in each plot) along with blue dots, each of which corresponds 
 to an extremal solution determined by the fixed value of $n$, namely  $n=0.71$ (left) and $n=0.70$ (right), and a particular value of $h_0$ in the corresponding admissible interval (c.f.  Fig.~\ref{fig:x0_curves}). The first thing to note is that the $n = 1/\sqrt{2}$ curve terminates at nonzero $(Q, J)$. This does not appear to be a precision issue but a general feature -- we have examined this to several hundred digits of precision. The second thing to note is that $n = 1/\sqrt{2}$ separates two distinct qualitative behaviors in the $(Q, J)$ plane, as we shall now describe.

Namely, when $n$ is larger than $1/(\sqrt{2} b)$, represented by left Figure \ref{fig:example_curves} for $n=0.71$,  
the blue `curve' begins at the point $(Q=0, J = n^2)$ when $h_0 = 0$. This curve hugs the $n = 1/\sqrt{2}$ curve, before ultimately splitting off and intersecting the horizontal axis at some finite $|Q| > 1/2$. Such  solutions, then, are continuously connected to the $J = 0$ extremal static solutions we considered earlier in this manuscript. (Recall, the existence of such solutions requires $|Q| > 1/2$ in units where $b = 1$.) All curves with $n > 1/\sqrt{2}$ follow this same pattern, and near $J = 0$ the small angular momentum results we derived in Section~\ref{sec:small_J} provide a good approximation to them. 

On the other hand, when $n$ is smaller than $1/(\sqrt{2} b)$, represented by right Figure~\ref{fig:example_curves} for $n=0.70$, the blue 
`curve' exhibits a qualitatively different behaviour. In this case, the curve begins hugging the inside of the $n = 1/\sqrt{2}$ curve but then shoots off inward, crossing $Q = 0$,  and oscillating back and forth several times, with the amplitude of oscillation decreasing with each successive crossing of the vertical axis. All curves with $n < 1/\sqrt{2}$ follow this same pattern. In fact, as we will discuss in more detail below,  the curves with $n < 1/\sqrt{2}$ exhibit a potentially infinite number of these oscillations and likely terminate at the $(Q=0, J=0)$ point. We shall support this by examining the asymptotic properties of our solutions.

\begin{figure}[t]
    \centering
    \includegraphics[width=0.75\linewidth]{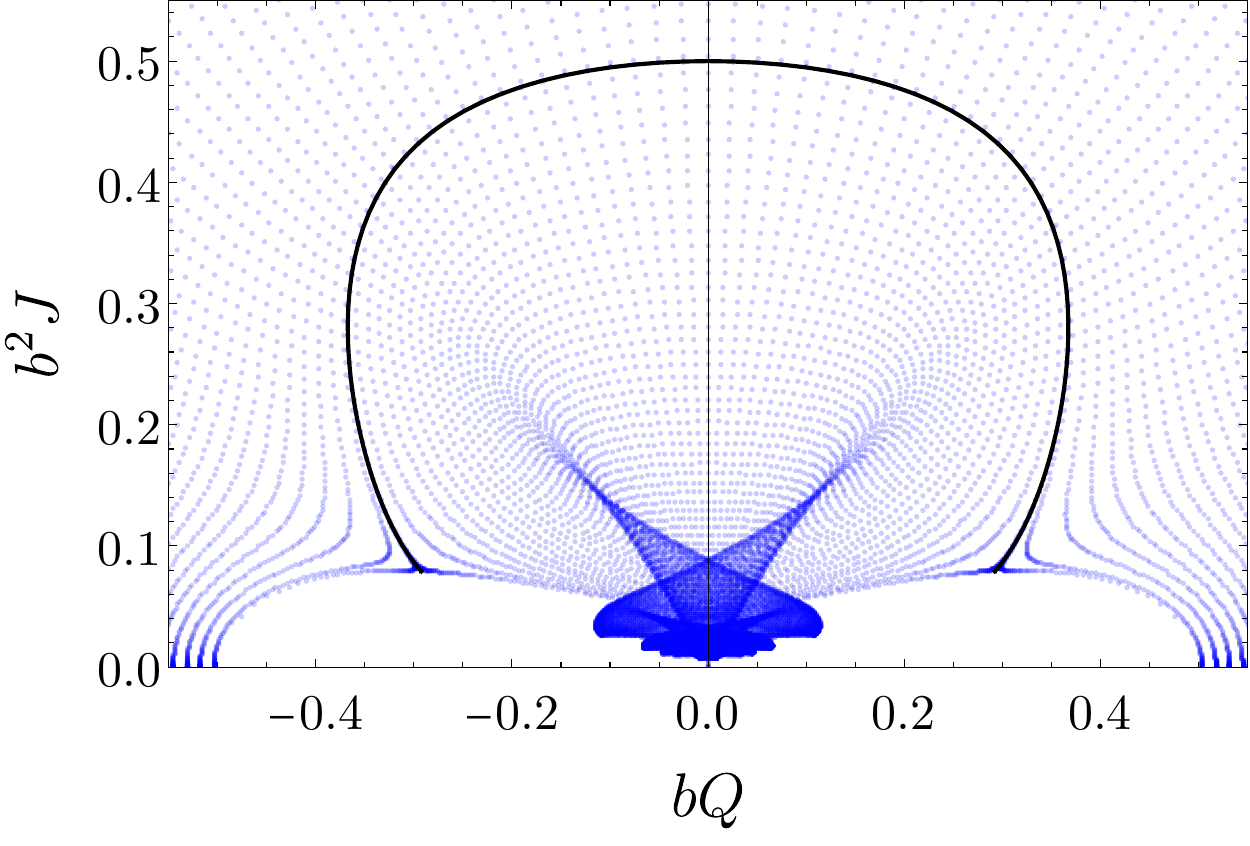}
    \caption{{\bf Phase space of extremal solutions in the $(Q, J)$ plane.} The black curve corresponds to $n = 1/\sqrt{2}$, while each blue dot is a solution for some choice of $(h_0, n)$.}
    \label{fig:phase_space}
\end{figure}

Having explained the qualitative behaviours, the general structure of the solution space can now be discussed. We show this in Figure~\ref{fig:phase_space}. In this plot, the black curve represents $n = 1/\sqrt{2}$, while each of the blue dots represent extremal solutions for particular $(n, h_0)$ values. To obtain the data used in this plot, we have fixed a particular value of $n$ and then selected 400 different values of $h_0$ within $h_0 \in [-h_{\rm max}, h_{\rm max}]$. To account for the sensitivity of $x_0$ near $\pm h_{\rm max}$, we have selected the data points according to a sampling function that weights in an exponential manner the data points near the endpoints of the interval. Specifically, we divide the interval $[-1, 1]$ into $N$ points $\{p_i\}$ and then take the $h_0$ data to be
\be 
(h_0)_i = h_{\rm max} \frac{\tanh(a p_i)}{\tanh{a}} \, .
\ee
Here $a$ is a weighting parameter, with larger $a$ yielding a stronger weighting of the points near the endpoint of the interval -- for Figure~\ref{fig:phase_space} we have set $a  = 10$. We have performed this for $n$ between $0.01$ and $1$ in steps of $0.01$, and in addition included data for $n = 1/\sqrt{2} \pm 10^{-4}$.

\begin{figure}[t]
    \centering
    \includegraphics[width=0.75\linewidth]{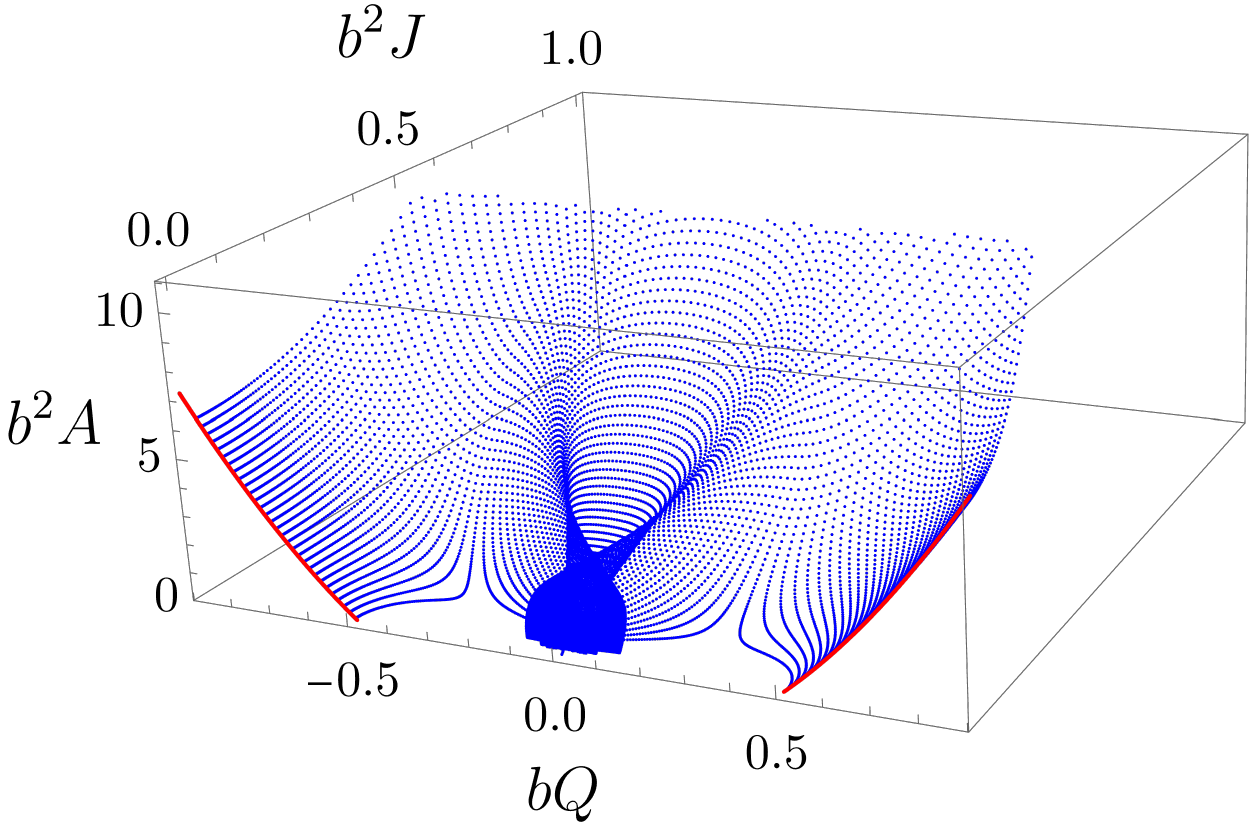}
    \caption{{\bf Extremal horizon area.} The area $A$ of extremal solutions is shown as a function of $(Q, J)$. Each blue dot corresponds to an extremal solution, while the red curves are the analytical result for the static solution when $J = 0$.}
    \label{fig:area_plot}
\end{figure}

There are two features of the plot that merit further elaboration. The first is the blue `butterfly'-like structure that exists near $J \approx 0$ and $Q \approx 0$. This structure arises simply due to the superposition of many of the oscillations that occur for each of the $n < 1/\sqrt{2}$ curves. While it has not been possible to capture \textit{all} such oscillations for each given $n$, we have captured those of the largest amplitude. That is, all our indications strongly suggest that the `butterfly' will not become any larger in size through additional refinement of the precision. 

The second and most striking feature of this plot are the regions of completely white space located between $-1/2 < Q < 1/2$ and $J \lesssim 0.08$. To help the reader identify the regions we are speaking of, the points $(Q, J) = (\pm 0.2, 0.01)$ are in these regions. Despite considerable effort, we have not been able to find regular near horizon extremal solutions in these regions of parameter space and suspect that none exist. We believe that these {\em white lobes} of the parameter space generalize the $|Q| > 1/2$ bound required for the existence of extremal static solutions  to the case of rotation. This is particularly exciting since we know that when $J = 0$ the full spacetime geometries for these parameter values have Schwarzschild-like causal structure~\cite{Hale:2025ezt}. That is, they are charged black holes without Cauchy horizons. It is very tempting to speculate that the same would be true for the parameter values within the white lobes of Figure~\ref{fig:phase_space}, and it would be very interesting to construct the full rotating spacetimes numerically and study their interior structure.

Let us continue to discuss the physical properties of the solution. In Figure~\ref{fig:area_plot} we show the area of the extremal horizon as a function of $(Q, J)$. In Einstein--Maxwell theory, the area function satisfied $A^2 = (8 \pi J)^2 + (4 \pi Q^2)^2$ and hence the analogous figure would have the appearance of a `spherical bowl'. In Einstein--Born--Infeld theory, provided both $Q$ and $J$ are large compared with $b$, the area begins to resemble the same `spherical bowl' structure, but it differs markedly for small $Q$ and $J$. Notably, near the boundary of the `white lobes' discussed from Figure~\ref{fig:phase_space} the area of the extremal horizon tends to zero, limiting to zero as $n \to 1/\sqrt{2}$. This adds further support to our earlier suggestion that no extremal solutions exist within the `white lobes' -- here we see that the boundary of that region corresponds to extremal solutions of \textit{vanishing area}. We also see that the function $A(Q, J)$ exhibits several self-intersections as a result of the `oscillating' behaviour of the constant $n$ curves for $n < 1/\sqrt{2}$. As a consistency check, we note that in the limit of $J = 0$, the area of the extremal horizon can be obtained analytically, \eqref{Area:static}, with 
$A=\pi [(2b Q)^2-1]/b^2$.
In Figure~\ref{fig:area_plot}, this case is included as the solid red lines -- we see that the numerical solutions with $n > 1/\sqrt{2}$ perfectly limit to this as $J \to 0$.

\begin{figure}
    \centering
    \includegraphics[width=0.75\linewidth]{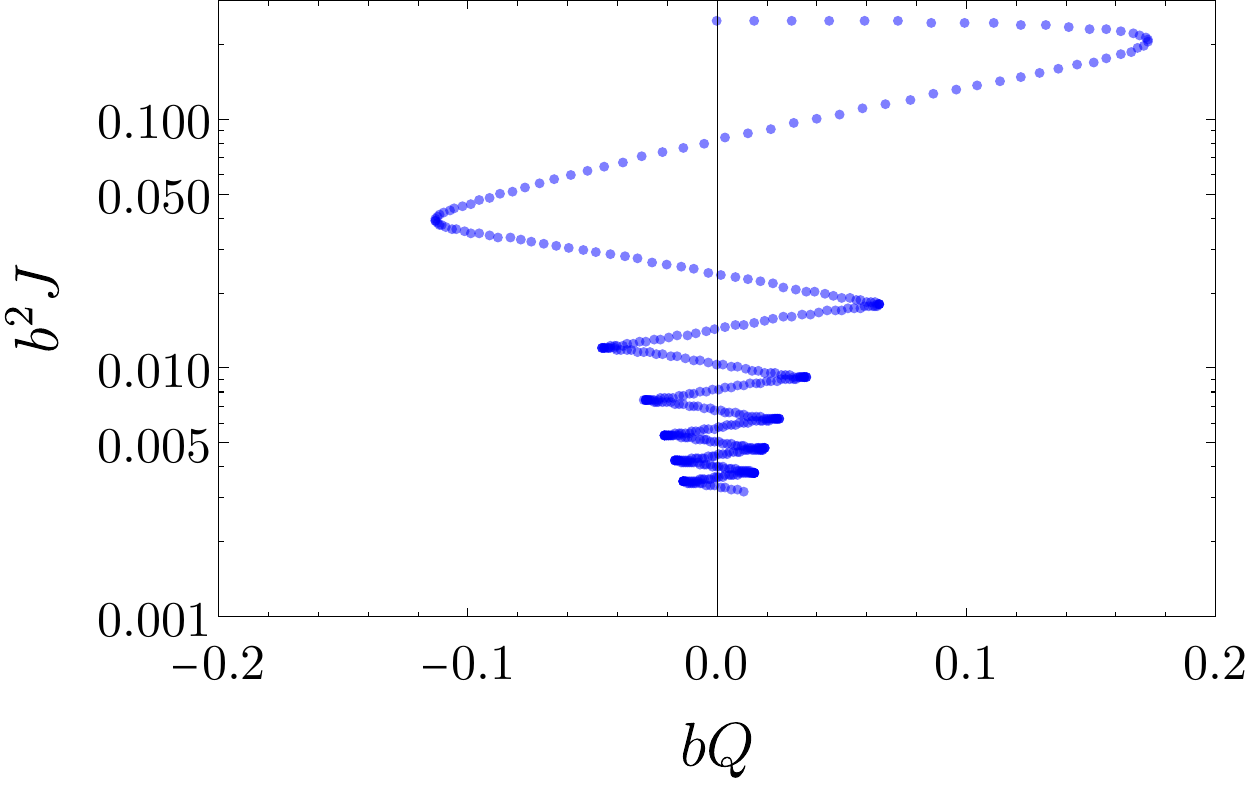}
    \caption{{\bf Oscillatory  feature.} Extremal solutions with $n = 0.5$ as a function of $h_0$ show in the $(Q, J)$ parameter space. Here we see 13 crossings axes, which requires working at very high precision (well above 100 digits in some cases). In this plot, we have included only the $h_0 < 0$ data for clarity; the $h_0 > 0$ is given simply by mapping $(Q, J) \to (-Q, J)$. }
    \label{fig:wiggly}
\end{figure}

Next, let us consider in more detail the {\em `oscillating' feature} mentioned earlier and consider its implications. All the curves with $n < 1/\sqrt{2}$ ultimately cross the vertical axis and then oscillate across the axis multiple times, with the amplitude of those oscillations (i.e. the extent to which the curves extend in the $\pm Q$ directions) decreasing with each successive crossing of the axis. In Figure~\ref{fig:wiggly} we show this feature for the example of $n = 0.5$, which in the figure has been tracked for 13 
axis crossings. The plot shows just `one half' of the $n = 0.5$ curve, with the values of $h_0$ ranging from $[-0.25, 0]$ in this case. We see clearly that the oscillations become `damped' in the charge direction, and compressed vertically in the angular momentum direction (hence why the plot uses a log scale).  In no case have we been able to numerically follow the oscillations to their final endpoint -- as discussed earlier, the precision required becomes severely limiting. However, we shall see in a moment that the asymptotics of our solutions allow us to draw robust conclusions. 

\begin{figure}
    \centering
    \includegraphics[width=0.95\linewidth]{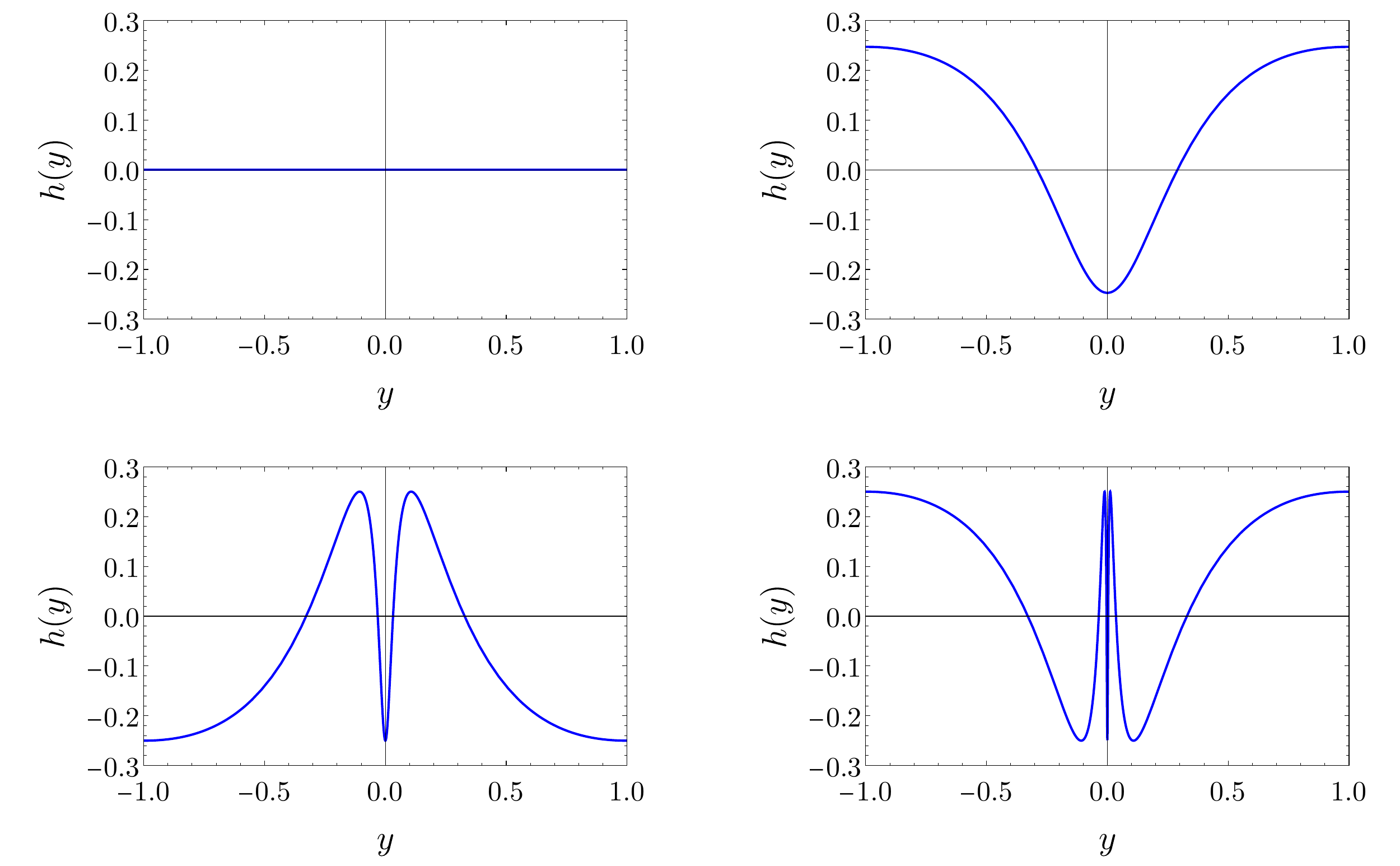}
    \includegraphics[width=0.45\linewidth]{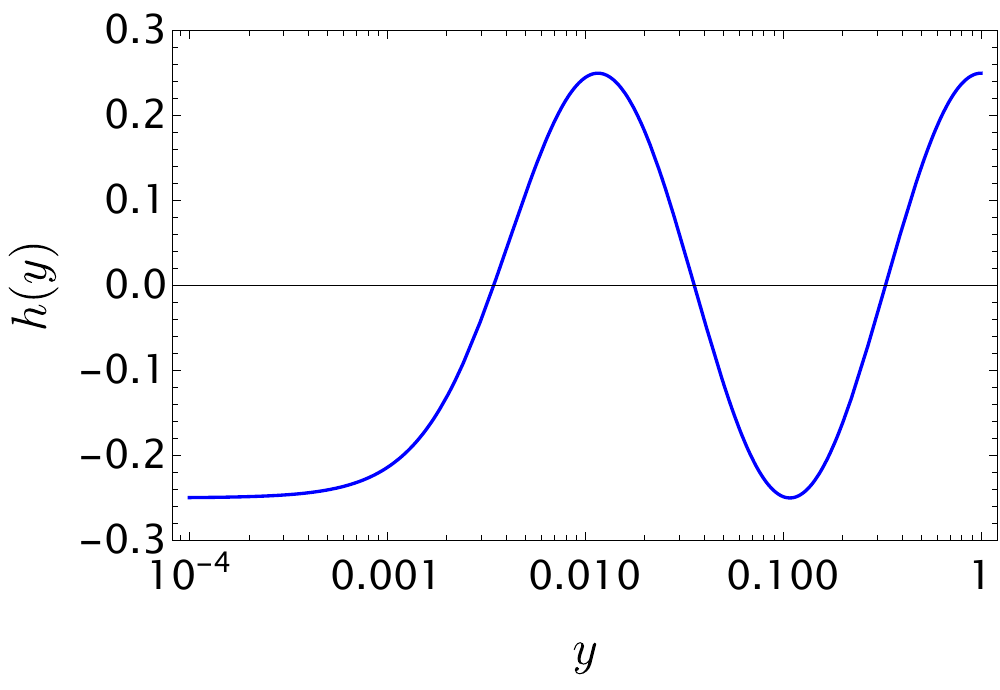}
    \caption{{\bf Configurations with vanishing electric charge}. Here we plot the vector potential function $h(y)$ for the first four zero charge configurations (crossings in Fig~\ref{fig:wiggly}) for  $n = 0.5$. In order from left-to-right top-to-bottom, the plots correspond to $h_0 = 0, -0.246957015, -0.249994317980, -0.249999974836$. We observe that with each successive crossing the function $h(y)$ ‘folds’ in on itself an
additional time. 
The last bottom figure, a `logarithmic magnification' of the previous figure for $h_0=-0.249999974836$,  clearly illustrates this feature.
}
    \label{fig:zero_charge}
\end{figure}

\subsubsection{Nontrivial field configurations with vanishing charge}

One interesting implication of the `oscillatory feature' illustrated in Figure~\ref{fig:wiggly} that we have not mentioned thus far is the following. Each time the curve crosses the vertical axis, we have \textit{vanishing charge}. These configurations represent near horizon extremal rotating black holes with vanishing charge but \textit{nonvanishing} electric and magnetic fields. This is truly an effect of the nonlinear nature of Born--Infeld theory, as in Maxwell theory the charge vanishes if and only if the fields also vanish. To further aid in understanding this feature, we plot in Figure~\ref{fig:zero_charge} the function $h(y)$ for the first four zero charge configurations with $n = 0.5$ (see also Figure~\ref{fig:zero_charge_gauge} for a confirmation that this is not a pure gauge). The first case (shown in the top-left plot) is the `trivial case' of vanishing field and vanishing charge, which occurs for $h_0 = 0$. The remaining three cases correspond to the first three non-trivial crossings. What we notice is that with each successive crossing the function $h(y)$ `folds' in on itself an additional time. For example, at the first crossing (top-right plot), the function $h(y)$ has a single local extremum. At the second crossing there are three local extrema, then five local extrema, with the suspected pattern of $2N+1$ extrema for $N$ crossings of the axis. 

We can study the `folding' property of the function $h(y)$ analytically by taking advantage of known {\em asymptotic} forms for the elliptic integrals appearing in our solution~\eqref{eq:exact_sol}. In terms of the notation used in the solution, the limit $|h_0| \to (bn)/2$ corresponds to $u \to 1/4$, or $m\to \infty$. In this limit, the function $h(y)$ has the following asymptotic form:
\be 
h(y) \sim \pm \frac{b n}{2} \cos \left(\sqrt{2} \log \frac{4 |T| \sqrt{m} }{1 + \sqrt{1+T^2}}\right) \qquad (m\to \infty)\,, 
\ee
where 
\be 
T = \frac{t}{\sqrt{1+2 \sqrt{u}}}\,,
\ee
and all other notation is identical to Eqs.~\eqref{eq:tu_def} and \eqref{eq:defns}. The argument of the cosine is accurate up to $1/m$ terms. Thus, the oscillations of the function we see in our numerical solution are clearly captured in the asymptotic form.

The asymptotics of the solution can actually allow us to say a bit more about the behaviour of the solution as $|h_0| \to (bn)/2$. We can obtain a similar asymptotic expansion of the function $g(y)$, but it is rather complicated and we do not present it here. However, from the asymptotic expansion of $g(y)$ we can obtain the asymptotic form of $x_0(n, h_0)$ valid as $|h_0| \to (bn)/2$,
\begin{align}
x_0 
&\sim \frac{1 - 2b^2 n^2}{\sqrt{2} \, n b^2 \log(m)} \qquad (m\to \infty) 
\nonumber\\
&= \frac{1 - 2b^2 n^2}{\sqrt{2} \, n b^2} 
   \cdot \left[\log \left( \frac{b n + 2 |h_0|}{b n - 2 |h_0|} \right) \right]^{-1}\,,
\end{align}
where in the second line we used the definition of $m$. This provides the leading asymptotic behaviour of $x_0$, making it clear that $x_0 \to 0$ as $|h_0| \to b n/2$. The logarithmic dependence appearing in here perfectly captures the rapid approach to $x_0 \to 0$ we saw and discussed in Figure~\ref{fig:x0_curves}. Moreover, note that we see that this asymptotic form yields a positive result only when $n^2 < 1/(2b^2)$, which is perfectly in line with our numerical observations made earlier.

\begin{figure}
    \centering
    \includegraphics[width=0.45\linewidth]{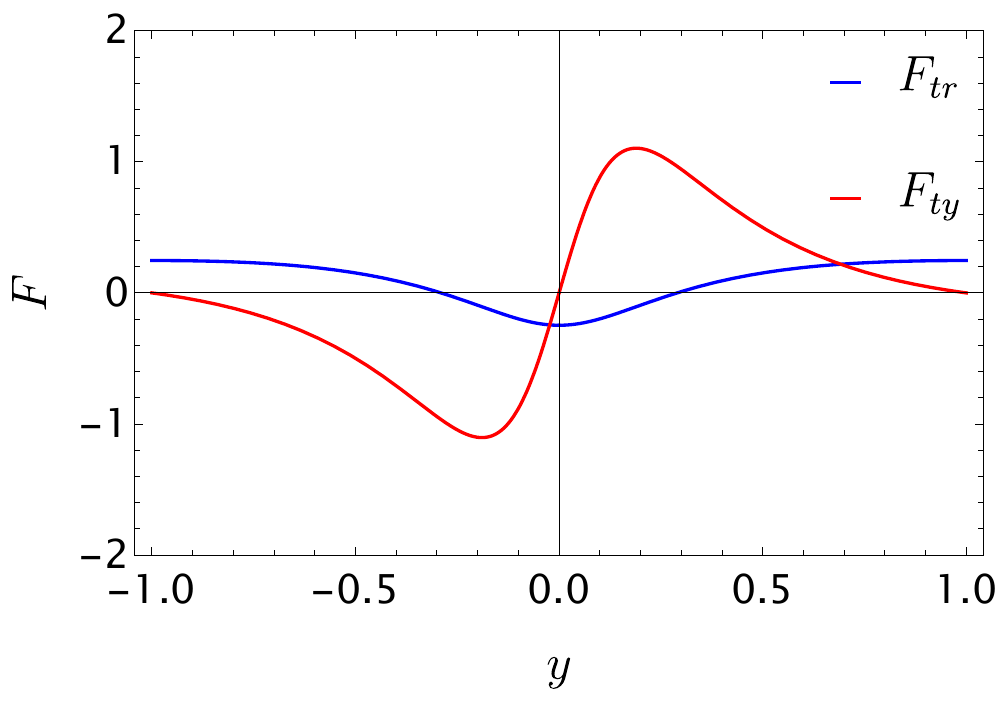}
    \includegraphics[width=0.45\linewidth]{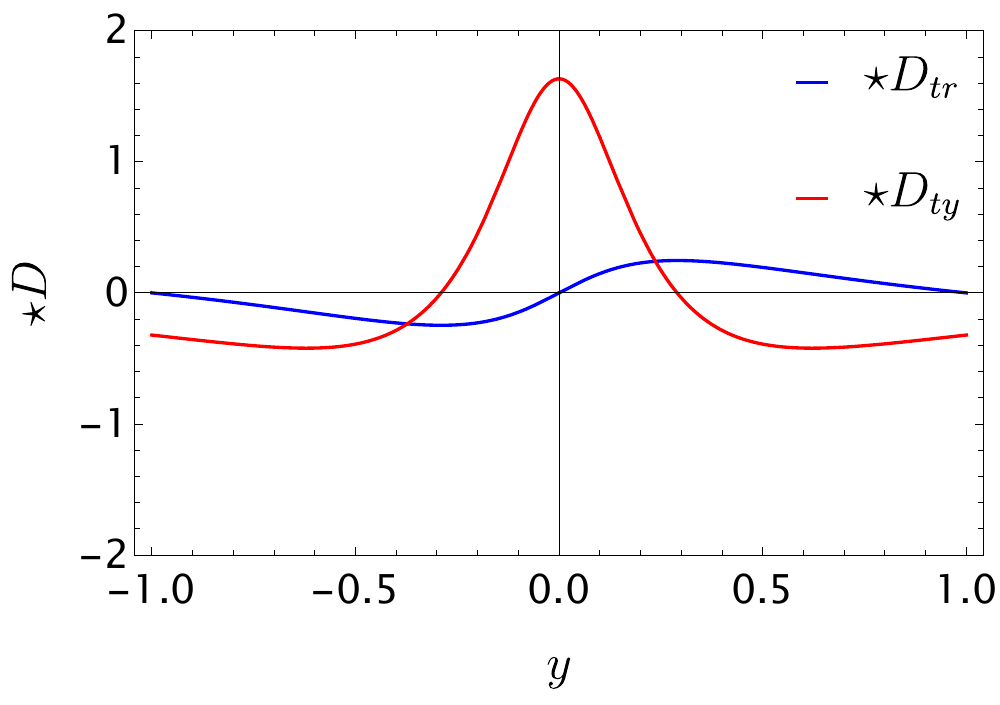}
    \caption{{\bf Configuration with vanishing electric charge: field strengths}. The two field `strengths' are displayed for $n = 1/2, b = 1$ and the configuration with  $h_0 = -0.246957015$. {\em Left.} Two non-trivial components of the electric field $F_{tr}$ and $F_{ty}$. {\em Right.} Two non-trivial components of the first row of the NLE-dual field $(\star D)_{tr}$ and $(\star D)_{ty}$.}
    \label{fig:zero_charge_gauge}
\end{figure}

\begin{figure}
    \centering
    \includegraphics[width=0.45\linewidth]{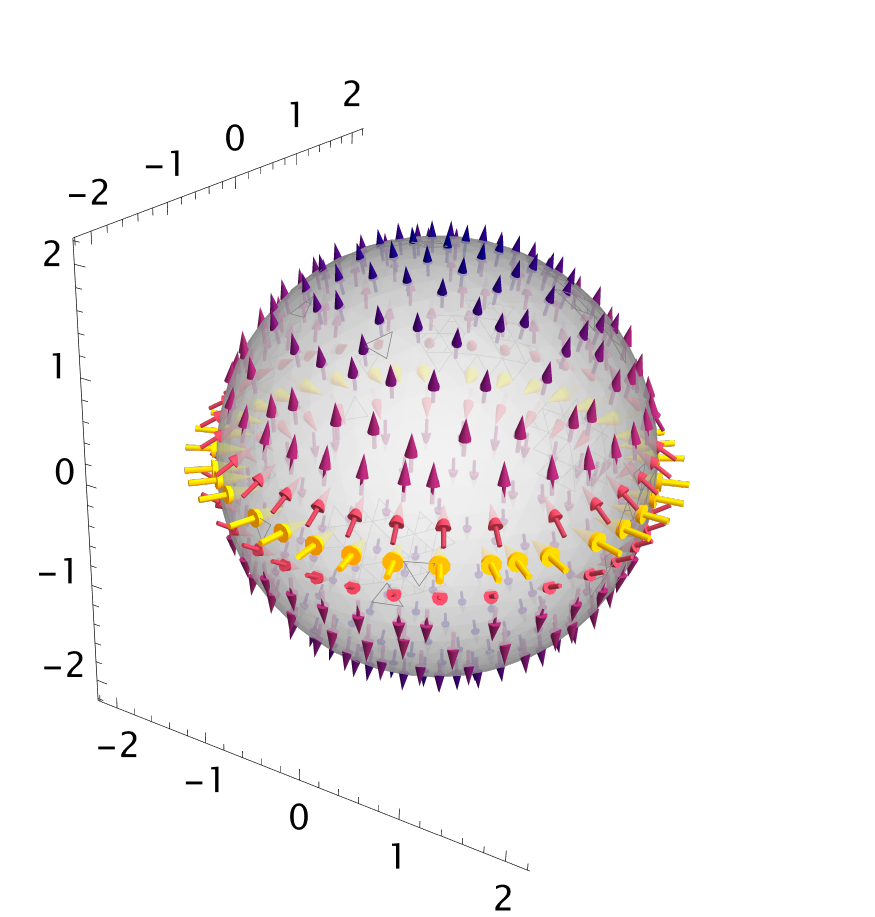}
    \caption{{\bf Configurations with vanishing electric charge: 3D visualization.} Here we display a visualisation of the electric part of the NLE-dual field $D$ projected onto the orthonormal tetrad corresponding to the metric \eqref{eq:ext_geom}, i.e., the vector field $(\left.e_{(r)}\right.^\mu D_{t\mu},\left.e_{(\theta)}\right.^\mu D_{t\mu},\left.e_{(\phi)}\right.^\mu D_{t\mu})$ for $n = 0.5, b = 1$, and $h_0 = -0.246957015$. }
    \label{fig:zero_charge_ball}
\end{figure}

In fact, we can use the asymptotic behaviour of the solution to derive the asymptotic form of the charge,
\be 
Q \sim \pm \frac{\left(2 b^2 n^2-1\right)}{{2 b^3 n^2 \log ^2(m)} } \sin \left(\sqrt{2} \sinh ^{-1}\left(\frac{\sqrt{m} \left(2 b^2
   n^2-1\right)}{2 b^2 n^2 \log (m)}\right)\right)\qquad (m\to \infty) \, ,
\ee
where the plus/minus corresponds to whether $h_0$ is positive or negative. This asymptotic form of the charge agrees with the qualitative features we observed in the numerics -- for large $m$, the charge oscillates with decreasing amplitude, ultimately tending to zero in the limit $m \to \infty$. Unfortunately, we have not been able to produce a useful asymptotic form for the angular momentum. 

Having now put on firm ground the asymptotic properties of the solutions in the strongly nonlinear regime, let us briefly return to the discussion of  
properties of the vanishing charge configurations. One thing that it is important to note is that while the charge vanishes, and hence the integral of $\star D$ vanishes, the quantity $\star D$ itself does not vanish pointwise, and the same is true for the field strength $F$, as we show in Figure~\ref{fig:zero_charge_gauge}. Hence it is a delicate cancellation under the integral that leads to the vanishing of the charge. If one considers that $\star D$ is akin to a local charge density, then we see from the right panel of the figure that there is a positive charge density concentrated around the equator, while approaching the poles reveals a slowly varying negative charge density. 
We provide a `three-dimensional visualization' of this field in Figure~\ref{fig:zero_charge_ball}.

\section{Discussion}
\label{Sec:6}

We have constructed extremal rotating black holes in Einstein--Born--Infeld theory in the near horizon regime. Remarkably, the solution to the Einstein--Born--Infeld equations can be written in a closed form in terms of elliptic integrals, which we studied using a combination of perturbative and numerical techniques. Using modified Komar integrals, we are able to study the charge and angular momentum of the extremal solutions and their relation to the black hole area/entropy, as well as verified the corresponding first law of near horizon mechanics. We now summarize the two most interesting results of our work.

\paragraph{Vanishing charge configurations.} We have observed that in the deeply nonlinear regime of Born--Infeld theory, there exist near horizon geometries which have vanishing charge but non-trivial electric/magnetic fields. This feature is absent in Maxwell electrodynamics and is a genuine consequence of nonlinear electrodynamics. We are not aware of other examples of this nature in the literature, and therefore our observation of this may be the first of its kind.\footnote{A similar, but apparently unrelated, feature arises for five-dimensional supersymmetric black rings. In that case, the near horizon geometry simply does not capture the charge~\cite{Kunduri:2013gce} -- see our Appendix~\ref{App:BTZ} for a related observation. Let us also note that this peculiarity is absent for the rotating black strings discussed in Appendix~\ref{App:strings}.} It would be interesting to better understand the physical implications of these solutions, and to assess whether or not the same feature can arise in simpler (e.g. flat spacetime) configurations in the absence of strong gravitational fields. 

This result also raises a puzzle. When the solution of interest has an asymptotically flat region, then at long distances from the source, Born--Infeld theory reduces to Maxwell and the asymptotic electric charge is the same in the two theories. In the near horizon geometries we have studied here, such an asymptotic region is lacking, and this is a crucial feature that underlies the result. What then happens when the near horizon geometries are `embedded' in full solutions with asymptotic regions? Since our analysis has been restricted to the near horizon region, we cannot answer this question here, but offer two {\em  speculations}. One possibility is that these particular zero charge configurations cannot be smoothly joined into a full spacetime geometry with an asymptotically flat region. In this case, they would be artifacts of the truncation to the near horizon region. If so, this would mean that a much larger portion of the parameter space does not support extremal black holes, calling into question what the structure of the full solutions would be in these cases. Another possibility, which is perhaps more intriguing, 
is that these zero charge configurations represent some form of `{\em discrete non-uniqueness}' of the zero charge configuration. From the full spacetime perspective, the particular electric/magnetic field configurations that give rise to zero charge would be `frozen' in the throat geometry and may not extend outside of this region. If this is the case, it would be interesting to understand from the full spacetime perspective whether such configurations require an extremal throat, or exist for  general rotating Einstein--Born--Infeld black holes, like those constructed numerically in~\cite{Cheng:2025kfz}. 

\paragraph{Bounds on the existence of extremal solutions.} Perhaps the most interesting finding of this work is that displayed 
in Figure~\ref{fig:phase_space}, which shows that there are regions of $(Q, J)$ parameter space for which extremal black holes do not exist in Einstein--Born--Infeld theory. In Einstein-Maxwell theory, extremal black holes exist in the full $(Q, J)$ plane, so this observation is a genuine novelty of nonlinear electrodynamics. 

We have suggested, based on analogy with the static limit of the solutions, that if the equations of motion were to be solved for values of $(Q, J)$ in these regions, then the full solutions may correspond to `Schwarzschild-like' rotating black holes, i.e. with spacelike singularity and no Cauchy horizon. What we know for certain is that if one begins at a point in the $(Q, J)$ plane for which extremal rotating black holes do not exist, and then send $J \to 0$, the result will necessarily be one of the static Schwarzschild branch solutions. Hence, by continuity we expect that the same causal structure may also apply to the rotating black holes in this excluded region of the $(Q, J)$ plane. It would be very interesting to construct the corresponding solutions numerically and examine their interior structure to see if this expectation is borne out. If so, these may be the first examples of stationary rotating black holes without Cauchy horizons.

\paragraph{Future directions.} We conclude here by mentioning several possibilities for future directions.

\begin{itemize}
    \item Here we have focused on Einstein--Born--Infeld theory, but we expect that much of the analysis can be extended to other theories of nonlinear electrodynamics. One drawback of Einstein--Born--Infeld theory is that the physical charges could not be integrated in closed form. It is conceivable that certain NLEs may exist that would circumvent this obstacle, and it would be interesting to identify them. For example, an analogous feature happens for certain specially selected higher derivative theories of gravity~\cite{Cano:2023dyg}. 
    \item We focused on the physically most interesting case of four spacetime dimensions and imposed asymptotic flatness and vanishing magnetic charge. It would nonetheless be interesting to consider near horizon geometries in higher dimensions. The equal-spinning rotating black holes in odd dimensions would be a particularly simple case to consider as the near horizon geometry is fixed up to a set of constants. 
    However, it can be shown that higher-dimensional static Einstein--Born--Infeld black holes, although admitting both S- and RN- branches of solutions, they  do not have a universal charge gap for the existence of extremal solutions. Thus, we do not expect this approach to shed light on the issue of whether or not higher-dimensional rotating black holes can lack Cauchy horizons. Also of interest could be the four-dimensional extensions of our work with magnetic charge and/or cosmological constant.  
    In particular, since with magnetic charge $h(y)$ is no longer an even function, the absence of a NUT charge would be slightly more subtle, and would likely require a nontrivial integration constant in the function $g(y)$. It should be possible to construct these solutions based on a duality rotation of those presented here~\cite{Murcia:2025psi}. The magnetic and dyonic solutions may lead to interesting new features worth exploring. 
    
    \item It would be particularly interesting and useful to perform a detailed numerical study of the four-dimensional rotating black holes of Einstein--Born--Infeld theory. This problem was very recently tackled in~\cite{Cheng:2025kfz}. However, those results begin to breakdown near extremality and, moreover, do not allow one to understand the black hole interior in this theory. As we have seen, the near horizon geometry allows for perturbative results for the area, charge, and angular momentum to be extracted, and it could be helpful to use these results as benchmarks in a complete numerical scheme. 
\end{itemize}

\section*{Acknowledgements}

We thank Marina David, Ivan Kol{\'a}{\v r}, James Lucietti and {\'A}ngel Murcia for helpful comments.  D.K. and T.H. are grateful for support from
GA\v{C}R 23-07457S grant of the Czech Science Foundation and the Charles University Research Center Grant
No. UNCE24/SCI/016.

\appendix



\section{Higher-order series coefficients}
\label{App:A}

\subsection{Extremal area}
Here we present several higher-order terms in the expression for the area of the extremal horizon, generalizing~\eqref{horizon_area}. Let us compress the notation as much as possible by introducing dimensionless charge and angular momentum parameters: $q \equiv b Q$ and $j \equiv b^2 J$. Then let us write the area as a function of the charge and angular momentum in the following schematic way:
\be 
A = \sum_i A^{(2i)} j^{2i}\,,
\ee
where $A^{(i)}$ are functions only of the charge. Then, we have the following results:
\begin{align}
    \frac{b^2}{4 \pi} A^{(0)} &=  q^2 - \frac{1}{4} \, ,
    \\
    \frac{b^2}{4 \pi} A^{(2)} &= \frac{24 \left(4 q^2+1\right)}{1+48 q^4} \, ,
    \\
    \frac{b^2}{4 \pi} A^{(4)} &=-\frac{1152}{5 \left(48 q^4+1\right)^4} \bigg(46080 q^{10}+29952 q^8-384 q^6+672 q^4-220 q^2-15\bigg) \, ,
    \\
    \frac{b^2}{4 \pi} A^{(6)} &= \frac{110592}{175 \left(48 q^4+1\right)^7} \bigg(3715891200 q^{18}+3583180800 q^{16}+3538944 q^{14}+69009408 q^{12}
    \nonumber
    \\
    &-99698688 q^{10}-7893504
   q^8-4392960 q^6-86400 q^4+37100 q^2+1575\bigg) \, ,
   \\
   \frac{b^2}{4 \pi} A^{(8)} &= -\frac{2654208}{875 \left(48 q^4+1\right)^{10}} \bigg(214035333120000 q^{26}+264690361958400 q^{24}
   \nonumber 
   \\
   &+12471804887040 q^{22}-3491323969536
   q^{20}-16905938927616 q^{18}
   \nonumber 
   \\
   &-2187656626176 q^{16}-1643072126976 q^{14}+41401073664
   q^{12}-28368092160 q^{10}
   \nonumber 
   \\
   &+6821648640 q^8+976245120 q^6+1471200 q^4-3762500 q^2-114625\bigg)  \, ,
   \\
   \frac{b^2}{4 \pi} A^{(10)} &= \frac{764411904}{336875 \left(48 q^4+1\right)^{13}} \bigg(88600354215690240000 q^{34}+131856723810975744000 q^{32}
   \nonumber 
   \\
   &+15020704513445068800
   q^{30}-9458777291678023680 q^{28}-15585086991199371264 q^{26}
   \nonumber 
   \\
   &-3206612103808942080
   q^{24}-2344644906225500160 q^{22}+119907001156239360 q^{20}
   \nonumber 
   \\
   &-109704331371479040
   q^{18}+34464369585291264 q^{16}-800110785331200 q^{14}
   \nonumber 
   \\
   &+1029208466718720 q^{12}+45463142215680
   q^{10}-13211939251200 q^8
   \nonumber 
   \\
   &-988240691200 q^6+5993344000 q^4+2507697500 q^2+59626875 \bigg) \, .
\end{align}
The first two coefficients are those given already in the main text. The complexity of the higher-order coefficients increases rapidly. 
Unfortunately, we have been unable to identify a pattern in these coefficients that would allow their resummation to be obtained in a closed form. 

\subsection{Charge for vanishing area}

Here we present additional terms appearing in the expansion of eq.~\eqref{area_charge}. We write 
\be 
b |Q_\star^{(J)}| = \sum_{i} Q^{(2i)} j^{2i} \, ,
\ee
where as before $j \equiv J b^2$ is the dimensionless angular momentum. Some additional coefficients not included in~\eqref{area_charge} are
\begin{align}
    Q^{(10)} &= -\frac{193247985973248}{336875} \, ,
    \\
    Q^{(12)} &= -\frac{1501505177924395008}{21896875} \, ,
    \\
    Q^{(14)} &= -\frac{6562735311612356001792}{766390625} \, ,
    \\
    Q^{(16)} &= -\frac{71914662431687661159776256}{65143203125} \, ,
    \\
    Q^{(18)} &= -\frac{69480234271721679557944319410176}{476522530859375} \, ,
    \\
    Q^{(20)} &= -\frac{1336173172503363782899337502130176}{68074647265625} \, ,
    \\
    Q^{(22)} &= -\frac{13380567057458480512129904769101885079552}{4986808285443359375} \, ,
    \\
    Q^{(24)} &= -\frac{9262330077732985750420862834240023443800064}{24934041427216796875} \, .
\end{align}

\section{Numerical verification of solution}
\label{app:numerical_check}

In this appendix, we consider solving the Einstein--Born--Infeld equations \textit{numerically}. This allows us to perform a manual `consistency check' for the closed form solution, which relies on certain built-in functions of Mathematica. 

Any approach to this problem has the following schematic form. The Born--Infeld equation is a second-order nonlinear differential equation, and hence can be specified by two free constants $h_0 = h(0)$ and $h_1 = h'(0)$. As we have discussed, $h_1$ is related to the magnetic charge and we therefore set it to zero, leaving one parameter coming from the electromagnetic equations. The gravitational equations have two specifiable parameters $(x_0, n)$ and one derived parameter $\omega$. We also demand the absence of the NUT charge by ensuring that $g'(0) = 0$. 

Each solution is then determined by three specifiable parameters $(h_0, x_0, n)$ which are, however, not independent but must be constrained by regularity; see eq.~\eqref{eq:regularity}. In our case, we can use the closed form expression for $g(y)$ obtained in Eq.~\eqref{eq:exact_sol} to quickly solve the regularity conditions. We specify values of $(n, h_0)$ and then use our exact form of $g(y)$ to obtain the correct value of $x_0$ by demanding that $g(1) = 0$. When this value of $x_0$ has been obtained, we use it as input for a fully numerical construction of $g(y)$ and $h(y)$. If the two methods are consistent, then the result should be a $g(y)$ and $h(y)$ that satisfy the regularity conditions and match the closed form solutions. 

We solve for $h(y)$ and $g(y)$ numerically in the following way. For the electromagnetic equation, we use Mathematica's \textsc{NDSolve}, supplying the conditions $h(0) = h_0$ and $h'(0) = 0$. The method we use for the Einstein equations is a bit more involved.  Due to the total derivative nature of eq.~\eqref{bi_efe}, the problem of finding $g(y)$ can be reduced to quadrature. In practice, one encounters a problem since the integral of the right-hand side of eq.~\eqref{bi_efe} exhibits a $1/y$ divergence at $y = 0$. This is of course canceled by the factor of $1/y$ that appears in the denominator of the left-hand side, but it is nonetheless problematic numerically. We can solve this issue by dealing with the problematic term at $y = 0$ by hand. To this end, let us define 
\be 
p(y) \equiv 8 \pi \left(y^2 x_0^2 + n^2 \right)^2 T_{yy} \, .
\ee
Then we can write a manifestly singularity-free integral for the function $g(y)$ in the following form:
\begin{align}
    g(y) &= \frac{(1-y) \left(y x_0^2 + n^2 \right)}{x_0^2 \left( y^2 x_0^2 + n^2 \right)} + \frac{(1-y) p(0)}{x_0^2 \left(y^2 x_0^2 + n^2 \right)} + \frac{y}{x_0^2 \left(y^2 x_0^2 + n^2 \right)} \int_{u = 1}^y \frac{p(u)-p(0)}{u^2} {\rm d}u \, .
\end{align}
Thus, upon plugging our numerically determined function $h(y)$ into this expression, we can obtain a numerical construction of $g(y)$ to compare with our exact result. 

\begin{figure}
    \centering
    \includegraphics[width=\linewidth]{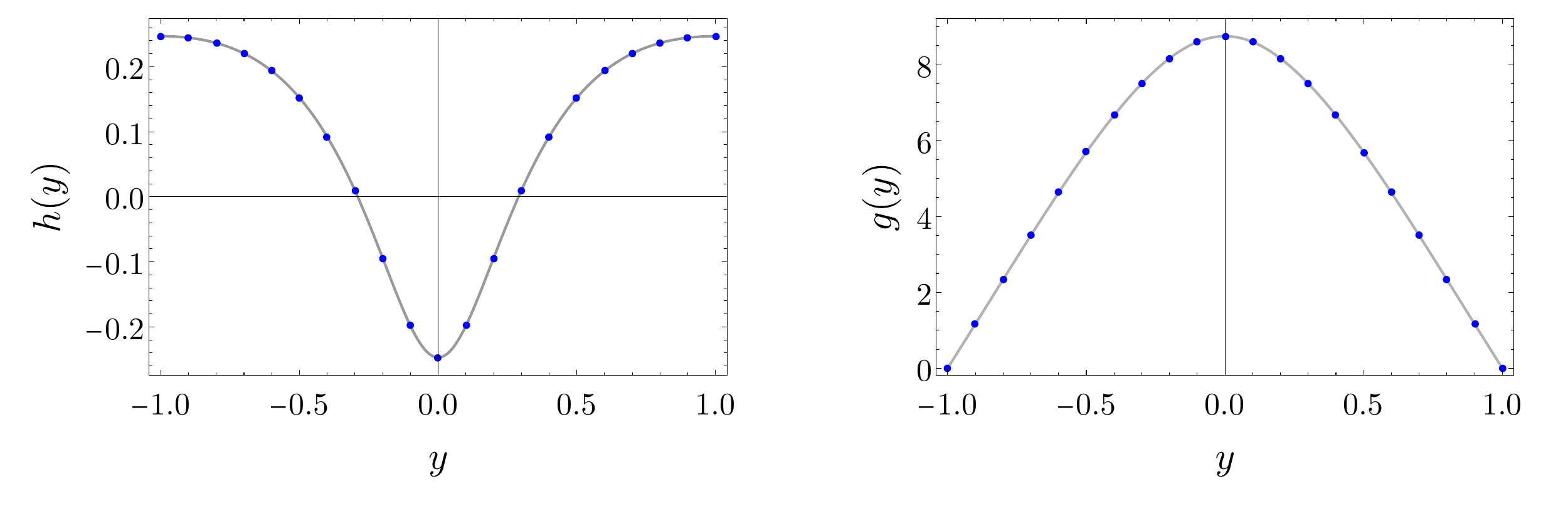}
    \caption{{\bf Numerical verification of solution}. Comparison between a fully numerical construction of the functions $h(y)$ and $g(y)$ with the closed form expressions showing perfect agreement. The blue dots are the numerical solution, while the gray curves are the closed-form results. The parameter choices here are $n = 1/2$, $h_0 = -0.246957015$ and $b = 1$. }
    \label{fig:num_anal_compare}
\end{figure}

We have performed robust tests of the parameter space numerically, fully reproducing (at a coarser resolution) the $(Q, J)$ solution space shown in Figure~\ref{fig:phase_space}. Here, we present in Figure~\ref{fig:num_anal_compare} a comparison between the numerical and closed form solutions for a particular set of parameters.

\section{Rotating Born--Infeld BTZ black holes}\label{App:BTZ}

In three dimensions, static charged black holes are known in any theory of NLE, e.g.,  \cite{Cataldo:2000ns}. Moreover,  such holes can be endowed with a `{\em topological rotation}' by employing a `boosting technique' \cite{Clement:1995zt, Martinez:1999qi} -- leading to rotating and NLE-charged BTZ black holes,  e.g., \cite{ Canate:2020btq}. In this appendix we shall review this construction for rotating Born--Infeld BTZ black holes and summarize their basic properties. 

The static Born--Infeld charged BTZ black hole reads \cite{Cataldo:1999wr, Myung:2008kd}\footnote{Note that in three dimensions the `determinant form' of the Born--Infeld Lagrangian reduces to 
\be 
\mathcal{L}_{\rm BI} = \frac{b^2}{4} \left(1 - \sqrt{1 + \frac{4\mathcal{S}}{b^2} } \right)\,,
\ee 
which is the form we employ for the calculations in this appendix.} 
\be\label{BTZtstatic}
\dd s^2=-f\dd t^2+\frac{\dd r^2}{f}+r^2 \dd\varphi^2\,,
\ee
with the metric function $f$ and the vector potential $A$ taking the following explicit form:
\ba\label{BIf}
f&=&\frac{r^2}{\ell^2}-m+\frac{1}{2}rb^2\Bigl(r-\sqrt{r^2+r_1^2}\Bigr)-\frac{1}{2}r_0b^2\Bigl(r_0-\sqrt{r_0^2+r_1^2}\Bigr)
-2e^2\ln \left(\frac{r+\sqrt{r^2+r_1^2}}{r_0+\sqrt{r_0^2+r_1^2}}\right)
\nonumber\\
&&\quad 
+e^2+\frac{r_0^2b^2}{2}\Bigl(1-\sqrt{1+r_1^2/r_0^2}\Bigr)+2e^2\ln \left(\frac{2}{1+\sqrt{1+r_1^2/r_0^2}}\right)
\,,\nonumber \\
A&=&\phi \dd t\,,\quad \phi=-e\ln \left(\frac{2(r+\sqrt{r^2+r_1^2})}{r_0\bigl(1+\sqrt{1+r_1^2/r_0^2}\bigr)}\right)\,,\quad r_1\equiv \frac{2e}{b}\,.
\ea 
Here, $\ell$ stands for the AdS radius, $e$ is related to the electric charge, $m$ is related to the mass of the hole, and $r_0>0$  is an arbitrary 
scale; various constant terms in the above guarantee that we recover the Maxwell behavior, namely $f_{\mbox{\tiny M}}=\frac{r^2}{\ell^2}-m-2e^2\ln(r/r_0)$ and $\phi_{\mbox{\tiny M}}=-e\ln(r/r_0)$, in both the expansion in $1/b^2$ and the expansion in $1/r$. Moreover, expanding $f$ near the origin, we recover\footnote{The self-energy in three dimensions is `IR divergent', behaving as $\ln(R/r_0)$ at large radius $R$. The $U_{\mbox{\tiny self}}$ we identify in the near-origin expansion of the metric function agrees with a simple regularization of the self-energy.} 
\be
f=-m+U_{\mbox{\tiny self}}-2b|e|r+O(r^2)\,,\quad U_{\mbox{\tiny self}}=2e^2\ln\bigl(r_0b/|e|\bigr)+e^2\,.
\ee
Thus the value of the metric function remains finite at the origin. It happens that we have a RN-type black hole (or possibly a naked singularity) for $U_{\mbox{\tiny self}}\geq m$ and an S-type solution for $U_{\mbox{\tiny self}}<m$. {Because of the logarithm, the self-energy contribution is not a monotonic function of charge, but has a maximum at $e = b r_0$. As a consequence, the S-type black holes arise for both weak and strong charge. Namely, for 
\be 
m > (br_0)^2
\ee 
the RN-type solutions cease to exist entirely leaving only the S-type ones -- this is displayed in Fig.~\ref{fig:BIBTZ}}.
In other words, upon fixing the scale $r_0$, we now have a universal upper bound on mass (and also charge) above which no Cauchy horizons can exist for static Born--Infeld BTZ black holes.

\begin{figure}[t]
    \centering
\includegraphics[width=0.45\linewidth]{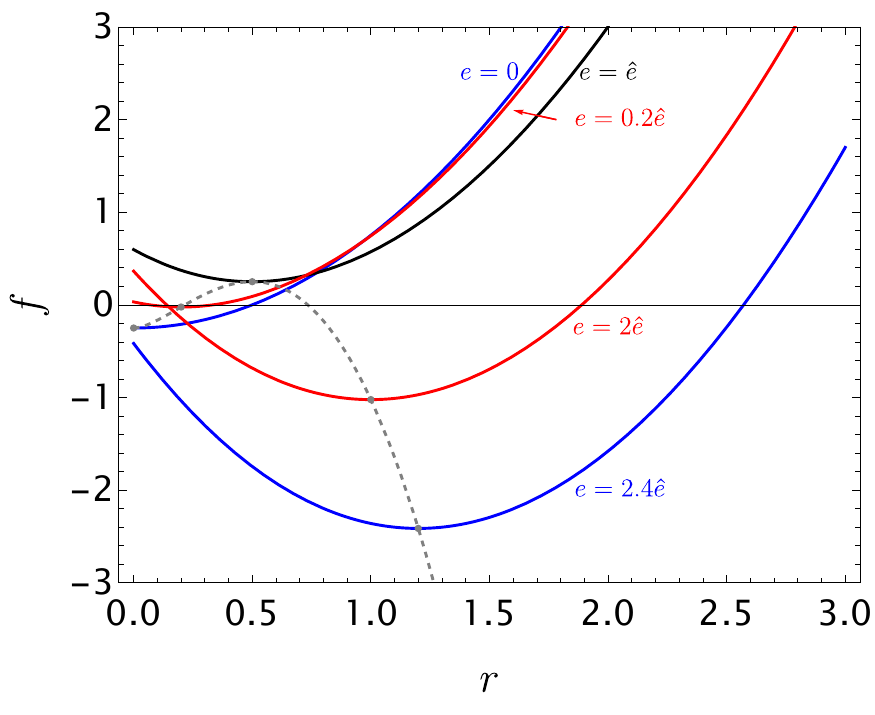}
\includegraphics[width=0.45\linewidth]{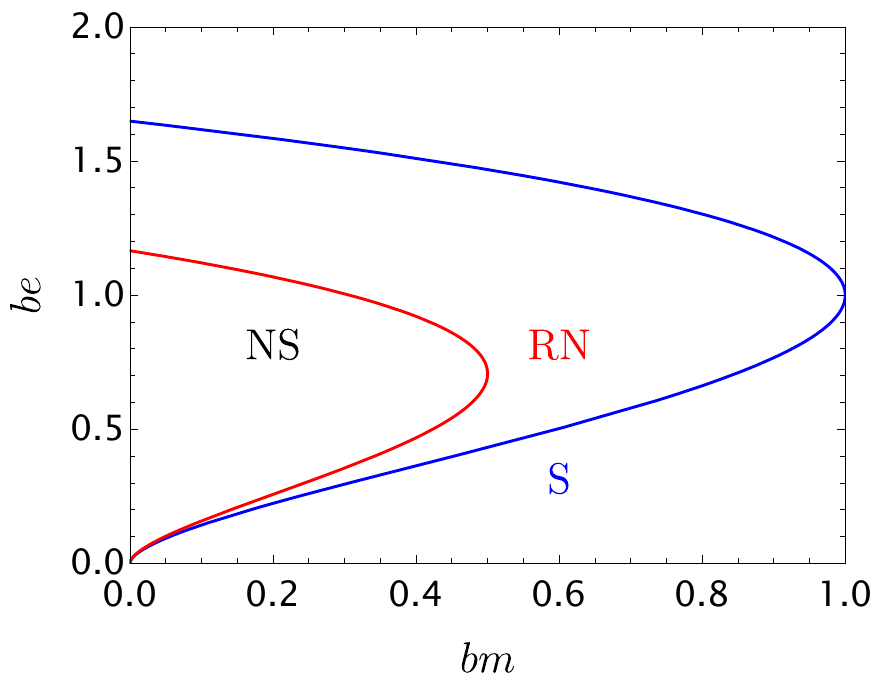}
    \caption{{\bf Static Born--Infeld BTZ horizons and parameter space}. {\em Left:} The metric function $f$ for the static Born--Infeld BTZ solution is displayed for fixed $m=0.25,\ell=1,r_0=1,b=1$ and for cases of the charge $e=0$ (top blue curve), $e=\hat{e}\equiv\frac{b r_0}{\sqrt{b^2 \ell^2+1}}$ (black curve), $e=2\hat{e}$ (red curve), and $e=2.4\hat{e}$ (bottom blue curve), where $\hat{e}$ corresponds to the peak of the grey dashed curve, which traces out the minimum of $f$ with varying $e$. {\em Right:} The $(bm,be)$ parameter space for static Born--Infeld BTZ for $\ell=1,r_0=1,b=1$. Upon fixing the scale $r_0$, we observe a universal upper bound on mass and charge above which no Cauchy horizons can exist; both the naked singularity (NS) and the RN-branch (RN) regions are now bounded -- surrounded by the S-branch (S) black holes.} 
    \label{fig:BIBTZ}
\end{figure}

To endow the above hole with rotation, we simply apply the following boost 
\cite{Lemos:1994xp, Lemos:1995cm}:
\be\label{tmf}
t \to \Xi t - a \varphi \, , \quad \varphi \to  \frac{at}{\ell^2} - \Xi \varphi \, ,\quad \Xi^2 = 1 + \frac{a^2}{\ell^2} \, ,
\ee
which yields the following solution:
\ba\label{btzfunny}
\dd s^2&=&-f(\Xi \dd t -a \dd\varphi)^2+\frac{r^2}{\ell^4}(a\dd t-\Xi \ell^2 \dd\varphi)^2+\frac{\dd r^2}{f}\,,\nonumber\\
A&=&\phi(\Xi \dd t-a \dd \varphi)\,.
\ea
Here, the `new' angular coordinate $\varphi$ is identified with period $2\pi$, $\varphi \sim \varphi+2\pi$, upon which the parameter $a$ is related to the rotation of the hole. The metric is written in Boyer--Lindquist-type coordinates which do not make manifest the AdS asymptotics of the solution. To bring the metric into a manifestly asymptotically AdS form, we must define a new `circumferential' radial coordinate $\hat{r}$ according to
\be 
\hat{r}^2 = r^2 \Xi^2 - f a^2 \, ,
\ee
which allows us to write the metric in the familiar ADM form,
\be 
{\rm d}s^2 = - N({\hat r}) F({\hat r}) {\rm d} t^2 + \hat{r}^2 \left({\rm d} \varphi - \frac{h({\hat r})}{2 \hat{r}^2} {\rm d}t\right)^2 + \frac{{\rm d} \hat{r}^2}{F({\hat r})}
\ee
where
\begin{align}
    N({\hat{r}}) &=\frac{4 r^2}{\left[2 r \Xi^2 - a^2 f'\right]^2} \, , \quad 
    F({\hat r}) = \frac{f}{4 \hat{r}^2} \left[2 r \Xi^2 - a^2 f'\right]^2 \, , \quad 
    h({\hat r}) = -2 a \Xi \left(f - \frac{r^2}{\ell^2} \right)\, ,
\end{align}
and $f'$ denotes the $r$ (not $\hat{r}$) derivative of the static metric function. It is understood that appearances of $r$ in this final result are to be replaced with the corresponding $r(\hat{r})$.  This relationship cannot in general be solved analytically, but for some applications (e.g. obtaining the charges) it is sufficient to invert it perturbatively at large radius.

The $\hat{r}_h$ values for which $F(\hat{r}_h) = 0$ give Killing horizons with associated null generator $\xi = \partial_t + \Omega_h \partial_\varphi$, where $\Omega_h = -h(\hat{r}_h)/(2 \hat{r}_h^2)$. As is clear from that equation, any horizon of the static solution, i.e.~where $f(r) = 0$, will also be a horizon of the rotating metric. Additional horizons can appear if the factor in square brackets vanishes in the limit $r = 0$ (to ensure the lapse remains finite). This is exactly what happens for the uncharged rotating BTZ metric. In that case, the inner horizon of the rotating hole corresponds to $r = 0$ in the unboosted coordinates. In the case of the rotating Maxwell-charged BTZ black hole, the bracketed term has no zeros and the horizon structure is completely determined by the static metric. As we have discussed above, the static Maxwell-charged BTZ black hole always has a Reissner--Nordstr{\"o}m-like causal structure, with an event and inner horizon. Hence, the rotating solution inherits this causal structure.

For the Born--Infeld case, there exists a region of parameter space for which the causal structure of the static solution is Schwarzschild-like. Moreover, we find no examples in which the bracketed term vanishes for finite charge and Born--Infeld coupling. Hence, in three-dimensions, Born--Infeld electrodynamics eliminates the Cauchy horizon of those rotating black holes that arise from boosted Schwarzschild-like solutions. We leave a more detailed analysis of these solutions to future work.

It is natural to wonder whether a near horizon extremal geometry could be used to extract bounds on the charge for the BIBTZ black holes. However, we find that the Born--Infeld equations force that the angular momentum or the charge must vanish -- near horizon extremal geometries with both charge and angular momentum do not exist. The same is true in three-dimensional Einstein--Maxwell--$\Lambda$ theory, as is well-known -- see the discussion on pages 38 and 39 of~\cite{Kunduri:2013gce}.

\section{Rotating black strings in Born--Infeld theory}\label{App:strings}
In this appendix we shall review the four-dimensional `cousins' of spherical rotating black holes studied in the main text -- rotating AdS black strings. Such solutions were constructed in \cite{Hendi:2010kv} (see also \cite{Hendi:2010zz}) and require the cosmological constant for their existence. 
Similar to spherical static black holes, they come in two (S-branch and RN-branch) types; with the former featuring no  Cauchy horizons.  
However, there is now no universal charge gap for the existence of RN-branch and extremal black holes with arbitrarily small charge may exist.

The solution takes the following form:
\ba 
\dd s^2&=&-f(\Xi \dd t-a \dd\varphi)^2+\frac{r^2}{\ell^4}(a\dd t-\Xi \ell^2 \dd\varphi)^2+\frac{\dd r^2}{f}+\frac{r^2}{\ell^2}\dd z^2\,,\nonumber\\
A&=&-\phi (\Xi \dd t-a \dd\varphi)\,,
\ea 
where $\ell$ corresponds to the AdS radius, $a$ is the rotation parameter, and $\Xi=\sqrt{1+a^2/\ell^2}$. For any NLE, the functions $f=f(r)$ and $\phi=\phi(r)$ are given by the corresponding black brane $(k=0)$ functions of the static AdS solution. In particular, for the Born--Infeld theory, we have 
\ba 
f_k&=&k-\frac{2M}{r}+\frac{r^2}{\ell^2}+\frac{2b^2}{r}\int_r^\infty\Bigl(\sqrt{r^4+\frac{Q^2}{b^2}}-r^2\Bigr)\dd r\label{D2}\nonumber\\
&=&
k-\frac{2M}{r}+\frac{r^2}{\ell^2}+\frac{2b^2r^2}{3}\Bigl(1-\sqrt{1+\frac{Q^2}{b^2 r^4}}\Bigr)+\frac{4Q^2}{3r^2} {}_2F_1\Bigl(\frac{1}{4},\frac{1}{2}; \frac{5}{4};-\frac{Q^2}{b^2r^4}\Bigr)\,,\\
\phi&=&\frac{Q}{r} {}_2F_1\Bigl(\frac{1}{4},\frac{1}{2}; \frac{5}{4};-\frac{Q^2}{b^2r^4}\Bigr)\,,\nonumber
\ea 
and identify $f\equiv f_0$. Note the `integration constant' associated with the upper limit in the integral for $f$ in \eqref{D2}. This ensures the proper asymptotic behavior on one side, as well `shifts' the overall `mass' by electromagnetic self-energy. Namely, we have the following expansion near the origin:
\be \label{C3}
f=-\frac{2(M-U_{\mbox{\tiny self}})}{r}-2b|Q|+O(r)\,,
\ee 
where the electromagnetic self energy reads
\be 
U_{\mbox{\tiny self}}=\frac{1}{6}\sqrt{\frac{b}{\pi}}|Q|^{3/2}\Gamma\Bigl(\frac{1}{4}\Bigr)^2\,.
\ee 
It is precisely this term that, similar to the static case, induces the possibility of having black holes without an inner Cauchy horizon, the so-called S-branch, e.g. \cite{Hale:2025ezt} (not discussed in \cite{Hendi:2010kv}). As detailed in \cite{Hale:2025ezt} the existence of this branch is a generic feature of NLE models with finite self-energy. 

Contrary to the static spherical case, however, there is now {\em no universal charge gap} for the existence of RN-branch black holes. Namely, as obvious from the expansion \eqref{C3}, for any non-trivial charge (no matter how small), the marginal $(M=U_{\mbox{\tiny self}})$ black holes feature negative $f$ in the origin, and thence there will exist non-trivial RN-branch black holes. We display the corresponding $(bM, bQ)$ parameter space for $\ell=1$ in Fig.~\ref{fig:Cstrings}, which is to be compared to Fig.~\ref{fig:BIbound} for spherical black holes where the universal bound is present.
\begin{figure}[t]
    \centering
\includegraphics[width=0.45\linewidth]{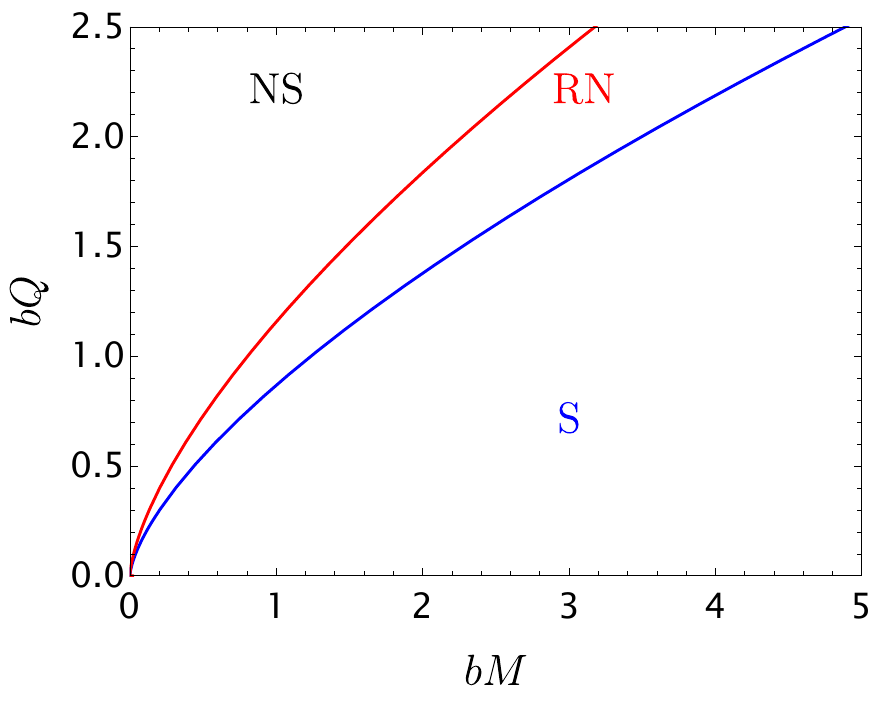}
    \caption{{\bf Black string parameter space}. The $(bM,bQ)$ parameter space of 4D Born-Infeld black strings for $\ell=1$. S-branch solutions exist below the blue curve. There is no charge or mass gap for the existence of the RN-branch (between red and blue curves). Naked singularity (NS) solutions exist above the red curve.}
    \label{fig:Cstrings}
\end{figure}

\bibliography{thebib}

\providecommand{\href}[2]{#2}\begingroup\raggedright\begin{thebibliography}{10}

\bibitem{Poisson:1989zz}
E.~Poisson and W.~Israel, \emph{{Inner-horizon instability and mass inflation in black holes}}, \href{http://dx.doi.org/10.1103/PhysRevLett.63.1663}{\emph{Phys. Rev. Lett.} {\bfseries 63} (1989) 1663--1666}.

\bibitem{Cardoso:2017soq}
V.~Cardoso, J.~L. Costa, K.~Destounis, P.~Hintz and A.~Jansen, \emph{{Quasinormal modes and Strong Cosmic Censorship}}, \href{http://dx.doi.org/10.1103/PhysRevLett.120.031103}{\emph{Phys. Rev. Lett.} {\bfseries 120} (2018) 031103}, [\href{https://arxiv.org/abs/1711.10502}{{\ttfamily 1711.10502}}].

\bibitem{Dias:2018etb}
O.~J.~C. Dias, H.~S. Reall and J.~E. Santos, \emph{{Strong cosmic censorship: taking the rough with the smooth}}, \href{http://dx.doi.org/10.1007/JHEP10(2018)001}{\emph{JHEP} {\bfseries 10} (2018) 001}, [\href{https://arxiv.org/abs/1808.02895}{{\ttfamily 1808.02895}}].

\bibitem{Hollands:2019whz}
S.~Hollands, R.~M. Wald and J.~Zahn, \emph{{Quantum instability of the Cauchy horizon in Reissner{\textendash}Nordstr{\"o}m{\textendash}deSitter spacetime}}, \href{http://dx.doi.org/10.1088/1361-6382/ab8052}{\emph{Class. Quant. Grav.} {\bfseries 37} (2020) 115009}, [\href{https://arxiv.org/abs/1912.06047}{{\ttfamily 1912.06047}}].

\bibitem{Hartnoll:2020rwq}
S.~A. Hartnoll, G.~T. Horowitz, J.~Kruthoff and J.~E. Santos, \emph{{Gravitational duals to the grand canonical ensemble abhor Cauchy horizons}}, \href{http://dx.doi.org/10.1007/JHEP10(2020)102}{\emph{JHEP} {\bfseries 10} (2020) 102}, [\href{https://arxiv.org/abs/2006.10056}{{\ttfamily 2006.10056}}].

\bibitem{Cai:2020wrp}
R.-G. Cai, L.~Li and R.-Q. Yang, \emph{{No Inner-Horizon Theorem for Black Holes with Charged Scalar Hairs}}, \href{http://dx.doi.org/10.1007/JHEP03(2021)263}{\emph{JHEP} {\bfseries 03} (2021) 263}, [\href{https://arxiv.org/abs/2009.05520}{{\ttfamily 2009.05520}}].

\bibitem{An:2021plu}
Y.-S. An, L.~Li and F.-G. Yang, \emph{{No Cauchy horizon theorem for nonlinear electrodynamics black holes with charged scalar hairs}}, \href{http://dx.doi.org/10.1103/PhysRevD.104.024040}{\emph{Phys. Rev. D} {\bfseries 104} (2021) 024040}, [\href{https://arxiv.org/abs/2106.01069}{{\ttfamily 2106.01069}}].

\bibitem{DeClerck:2023fax}
M.~De~Clerck, S.~A. Hartnoll and J.~E. Santos, \emph{{Mixmaster chaos in an AdS black hole interior}}, \href{http://dx.doi.org/10.1007/JHEP07(2024)202}{\emph{JHEP} {\bfseries 07} (2024) 202}, [\href{https://arxiv.org/abs/2312.11622}{{\ttfamily 2312.11622}}].

\bibitem{Born:1934gh}
M.~Born and L.~Infeld, \emph{{Foundations of the new field theory}}, \href{http://dx.doi.org/10.1098/rspa.1934.0059}{\emph{Proc. Roy. Soc. Lond. A} {\bfseries 144} (1934) 425--451}.

\bibitem{Plebanski:1970zz}
J.~Plebanski, \emph{{LECTURES ON NON LINEAR ELECTRODYNAMICS}}, .

\bibitem{Bialynicki-Birula:1992rcm}
I.~Bialynicki-Birula, \emph{{Field theory of photon dust}}, {\emph{Acta Phys. Polon. B} {\bfseries 23} (1992) 553--559}.

\bibitem{Gibbons:1995cv}
G.~W. Gibbons and D.~A. Rasheed, \emph{{Electric - magnetic duality rotations in nonlinear electrodynamics}}, \href{http://dx.doi.org/10.1016/0550-3213(95)00409-L}{\emph{Nucl. Phys. B} {\bfseries 454} (1995) 185--206}, [\href{https://arxiv.org/abs/hep-th/9506035}{{\ttfamily hep-th/9506035}}].

\bibitem{Russo:2024xnh}
J.~G. Russo and P.~K. Townsend, \emph{{Causality and energy conditions in nonlinear electrodynamics}}, \href{http://dx.doi.org/10.1007/JHEP06(2024)191}{\emph{JHEP} {\bfseries 06} (2024) 191}, [\href{https://arxiv.org/abs/2404.09994}{{\ttfamily 2404.09994}}].

\bibitem{Fradkin:1985qd}
E.~S. Fradkin and A.~A. Tseytlin, \emph{{Nonlinear Electrodynamics from Quantized Strings}}, \href{http://dx.doi.org/10.1016/0370-2693(85)90205-9}{\emph{Phys. Lett. B} {\bfseries 163} (1985) 123--130}.

\bibitem{Leigh:1989jq}
R.~G. Leigh, \emph{{Dirac-Born-Infeld Action from Dirichlet Sigma Model}}, \href{http://dx.doi.org/10.1142/S0217732389003099}{\emph{Mod. Phys. Lett. A} {\bfseries 4} (1989) 2767}.

\bibitem{Hale:2025ezt}
T.~Hale, R.~A. Hennigar and D.~Kubiznak, \emph{{Excising Cauchy Horizons with Nonlinear Electrodynamics}},  \href{https://arxiv.org/abs/2506.20802}{{\ttfamily 2506.20802}}.

\bibitem{Allahverdizadeh:2013oha}
M.~Allahverdizadeh, J.~P.~S. Lemos and A.~Sheykhi, \emph{{Extremal Myers-Perry black holes coupled to Born-Infeld electrodynamics in five dimensions}}, \href{http://dx.doi.org/10.1103/PhysRevD.87.084002}{\emph{Phys. Rev. D} {\bfseries 87} (2013) 084002}, [\href{https://arxiv.org/abs/1302.5079}{{\ttfamily 1302.5079}}].

\bibitem{Allaverdizadeh:2013rua}
M.~Allaverdizadeh, S.~H. Hendi, J.~P.~S. Lemos and A.~Sheykhi, \emph{{Extremal Myers-Perry black holes coupled to Born-Infeld electrodynamics in odd dimensions}}, \href{http://dx.doi.org/10.1142/S0218271814500321}{\emph{Int. J. Mod. Phys. D} {\bfseries 23} (2014) 1450032}, [\href{https://arxiv.org/abs/1304.0836}{{\ttfamily 1304.0836}}].

\bibitem{Kubiznak:2022vft}
D.~Kubiznak, T.~Tahamtan and O.~Svitek, \emph{{Slowly rotating black holes in nonlinear electrodynamics}}, \href{http://dx.doi.org/10.1103/PhysRevD.105.104064}{\emph{Phys. Rev. D} {\bfseries 105} (2022) 104064}, [\href{https://arxiv.org/abs/2203.01919}{{\ttfamily 2203.01919}}].

\bibitem{Hendi:2010kv}
S.~H. Hendi, \emph{{Rotating Black String with Nonlinear Source}}, \href{http://dx.doi.org/10.1103/PhysRevD.82.064040}{\emph{Phys. Rev. D} {\bfseries 82} (2010) 064040}, [\href{https://arxiv.org/abs/1008.5210}{{\ttfamily 1008.5210}}].

\bibitem{Hendi:2010zz}
S.~H. Hendi, \emph{{Rotating black branes in the presence of nonlinear electromagnetic field}}, \href{http://dx.doi.org/10.1140/epjc/s10052-010-1359-6}{\emph{Eur. Phys. J. C} {\bfseries 69} (2010) 281--288}, [\href{https://arxiv.org/abs/1008.0168}{{\ttfamily 1008.0168}}].

\bibitem{Cheng:2025kfz}
L.~Cheng and P.~Wang, \emph{{Rotating Black Holes in Einstein-Born-Infeld Theory}},  \href{https://arxiv.org/abs/2507.00879}{{\ttfamily 2507.00879}}.

\bibitem{Kunduri:2007vf}
H.~K. Kunduri, J.~Lucietti and H.~S. Reall, \emph{{Near-horizon symmetries of extremal black holes}}, \href{http://dx.doi.org/10.1088/0264-9381/24/16/012}{\emph{Class. Quant. Grav.} {\bfseries 24} (2007) 4169--4190}, [\href{https://arxiv.org/abs/0705.4214}{{\ttfamily 0705.4214}}].

\bibitem{Kunduri:2013gce}
H.~K. Kunduri and J.~Lucietti, \emph{{Classification of near-horizon geometries of extremal black holes}}, \href{http://dx.doi.org/10.12942/lrr-2013-8}{\emph{Living Rev. Rel.} {\bfseries 16} (2013) 8}, [\href{https://arxiv.org/abs/1306.2517}{{\ttfamily 1306.2517}}].

\bibitem{Astefanesei:2006dd}
D.~Astefanesei, K.~Goldstein, R.~P. Jena, A.~Sen and S.~P. Trivedi, \emph{{Rotating attractors}}, \href{http://dx.doi.org/10.1088/1126-6708/2006/10/058}{\emph{JHEP} {\bfseries 10} (2006) 058}, [\href{https://arxiv.org/abs/hep-th/0606244}{{\ttfamily hep-th/0606244}}].

\bibitem{Cano:2019ozf}
P.~A. Cano and D.~Pere{\~n}iguez, \emph{{Extremal Rotating Black Holes in Einsteinian Cubic Gravity}}, \href{http://dx.doi.org/10.1103/PhysRevD.101.044016}{\emph{Phys. Rev. D} {\bfseries 101} (2020) 044016}, [\href{https://arxiv.org/abs/1910.10721}{{\ttfamily 1910.10721}}].

\bibitem{plebanski1970lectures}
J.~Pleba{\'n}ski, \emph{Lectures on non-linear electrodynamics: an extended version of lectures given at the Niels Bohr Institute and NORDITA, Copenhagen, in October 1968}, vol.~25.
\newblock Nordita, 1970.

\bibitem{Hanaki:2007mb}
K.~Hanaki, K.~Ohashi and Y.~Tachikawa, \emph{{Comments on charges and near-horizon data of black rings}}, \href{http://dx.doi.org/10.1088/1126-6708/2007/12/057}{\emph{JHEP} {\bfseries 12} (2007) 057}, [\href{https://arxiv.org/abs/0704.1819}{{\ttfamily 0704.1819}}].

\bibitem{GarciaD:1984xrg}
A.~Garc{\'\i}a~D., H.~Salazar~I. and J.~F. Pleba{\'n}ski, \emph{{Type-D solutions of the Einstein and Born-Infeld nonlinear-electrodynamics equations}}, \href{http://dx.doi.org/10.1007/BF02721649}{\emph{Nuovo Cim. B} {\bfseries 84} (1984) 65--90}.

\bibitem{deOliveira:1994in}
H.~P. de~Oliveira, \emph{{Nonlinear charged black holes}}, \href{http://dx.doi.org/10.1088/0264-9381/11/6/012}{\emph{Class. Quant. Grav.} {\bfseries 11} (1994) 1469--1482}.

\bibitem{Fernando:2003tz}
S.~Fernando and D.~Krug, \emph{{Charged black hole solutions in Einstein-Born-Infeld gravity with a cosmological constant}}, \href{http://dx.doi.org/10.1023/A:1021315214180}{\emph{Gen. Rel. Grav.} {\bfseries 35} (2003) 129--137}, [\href{https://arxiv.org/abs/hep-th/0306120}{{\ttfamily hep-th/0306120}}].

\bibitem{Dey:2004yt}
T.~K. Dey, \emph{{Born-Infeld black holes in the presence of a cosmological constant}}, \href{http://dx.doi.org/10.1016/j.physletb.2004.06.047}{\emph{Phys. Lett. B} {\bfseries 595} (2004) 484--490}, [\href{https://arxiv.org/abs/hep-th/0406169}{{\ttfamily hep-th/0406169}}].

\bibitem{Cai:2004eh}
R.-G. Cai, D.-W. Pang and A.~Wang, \emph{{Born-Infeld black holes in (A)dS spaces}}, \href{http://dx.doi.org/10.1103/PhysRevD.70.124034}{\emph{Phys. Rev. D} {\bfseries 70} (2004) 124034}, [\href{https://arxiv.org/abs/hep-th/0410158}{{\ttfamily hep-th/0410158}}].

\bibitem{Hoffmann:1935ty}
B.~Hoffmann, \emph{{Gravitational and electromagnetic mass in the Born-Infeld electrodynamics}}, \href{http://dx.doi.org/10.1103/PhysRev.47.877}{\emph{Phys. Rev.} {\bfseries 47} (1935) 877--880}.

\bibitem{Sokolov:2025ara}
V.~Sokolov, \emph{{Extreme black points in Born-Infeld electrodynamics}},  \href{https://arxiv.org/abs/2509.00477}{{\ttfamily 2509.00477}}.

\bibitem{Sokolov:2025vtl}
V.~Sokolov, \emph{{Thermodynamics of charged black points in vacuum nonlinear electrodynamics}},  \href{https://arxiv.org/abs/2509.00473}{{\ttfamily 2509.00473}}.

\bibitem{Hajian:2013lna}
K.~Hajian, A.~Seraj and M.~M. Sheikh-Jabbari, \emph{{NHEG Mechanics: Laws of Near Horizon Extremal Geometry (Thermo)Dynamics}}, \href{http://dx.doi.org/10.1007/JHEP03(2014)014}{\emph{JHEP} {\bfseries 03} (2014) 014}, [\href{https://arxiv.org/abs/1310.3727}{{\ttfamily 1310.3727}}].

\bibitem{Hajian:2014twa}
K.~Hajian, A.~Seraj and M.~M. Sheikh-Jabbari, \emph{{Near Horizon Extremal Geometry Perturbations: Dynamical Field Perturbations vs. Parametric Variations}}, \href{http://dx.doi.org/10.1007/JHEP10(2014)111}{\emph{JHEP} {\bfseries 10} (2014) 111}, [\href{https://arxiv.org/abs/1407.1992}{{\ttfamily 1407.1992}}].

\bibitem{Newman:1965my}
E.~T. Newman, E.~Couch, K.~Chinnapared, A.~Exton, A.~Prakash and R.~Torrence, \emph{{Metric of a Rotating, Charged Mass}}, \href{http://dx.doi.org/10.1063/1.1704351}{\emph{J. Math. Phys.} {\bfseries 6} (1965) 918--919}.

\bibitem{Newman:1965tw}
E.~T. Newman and A.~I. Janis, \emph{{Note on the Kerr spinning particle metric}}, \href{http://dx.doi.org/10.1063/1.1704350}{\emph{J. Math. Phys.} {\bfseries 6} (1965) 915--917}.

\bibitem{Breton:2014mba}
N.~Bret{\'o}n and C.~E. Ram{\'\i}rez-Codiz, \emph{{On the NUT{\textendash}Born{\textendash}Infeld-{\ensuremath{\Lambda}} spacetime}}, \href{http://dx.doi.org/10.1016/j.aop.2014.11.016}{\emph{Annals Phys.} {\bfseries 353} (2014) 252--270}, [\href{https://arxiv.org/abs/1408.4376}{{\ttfamily 1408.4376}}].

\bibitem{BallonBordo:2020jtw}
A.~Ballon~Bordo, D.~Kubiz{\v{n}}{\'a}k and T.~R. Perche, \emph{{Taub-NUT solutions in conformal electrodynamics}}, \href{http://dx.doi.org/10.1016/j.physletb.2021.136312}{\emph{Phys. Lett. B} {\bfseries 817} (2021) 136312}, [\href{https://arxiv.org/abs/2011.13398}{{\ttfamily 2011.13398}}].

\bibitem{Cassani:2023vsa}
D.~Cassani, A.~Ruip{\'e}rez and E.~Turetta, \emph{{Boundary terms and conserved charges in higher-derivative gauged supergravity}}, \href{http://dx.doi.org/10.1007/JHEP06(2023)203}{\emph{JHEP} {\bfseries 06} (2023) 203}, [\href{https://arxiv.org/abs/2304.06101}{{\ttfamily 2304.06101}}].

\bibitem{Cano:2024tcr}
P.~A. Cano and M.~David, \emph{{Near-horizon geometries and black hole thermodynamics in higher-derivative AdS$_{5}$ supergravity}}, \href{http://dx.doi.org/10.1007/JHEP03(2024)036}{\emph{JHEP} {\bfseries 03} (2024) 036}, [\href{https://arxiv.org/abs/2402.02215}{{\ttfamily 2402.02215}}].

\bibitem{Cano:2023dyg}
P.~A. Cano and M.~David, \emph{{The extremal Kerr entropy in higher-derivative gravities}}, \href{http://dx.doi.org/10.1007/JHEP05(2023)219}{\emph{JHEP} {\bfseries 05} (2023) 219}, [\href{https://arxiv.org/abs/2303.13286}{{\ttfamily 2303.13286}}].

\bibitem{Murcia:2025psi}
{\'A}.~J. Murcia, \emph{{Novel duality-invariant theories of electrodynamics}},  \href{https://arxiv.org/abs/2507.16502}{{\ttfamily 2507.16502}}.

\bibitem{Cataldo:2000ns}
M.~Cataldo and A.~Garcia, \emph{{Regular (2+1)-dimensional black holes within nonlinear electrodynamics}}, \href{http://dx.doi.org/10.1103/PhysRevD.61.084003}{\emph{Phys. Rev. D} {\bfseries 61} (2000) 084003}, [\href{https://arxiv.org/abs/hep-th/0004177}{{\ttfamily hep-th/0004177}}].

\bibitem{Clement:1995zt}
G.~Clement, \emph{{Spinning charged BTZ black holes and selfdual particle - like solutions}}, \href{http://dx.doi.org/10.1016/0370-2693(95)01464-0}{\emph{Phys. Lett. B} {\bfseries 367} (1996) 70--74}, [\href{https://arxiv.org/abs/gr-qc/9510025}{{\ttfamily gr-qc/9510025}}].

\bibitem{Martinez:1999qi}
C.~Martinez, C.~Teitelboim and J.~Zanelli, \emph{{Charged rotating black hole in three space-time dimensions}}, \href{http://dx.doi.org/10.1103/PhysRevD.61.104013}{\emph{Phys. Rev. D} {\bfseries 61} (2000) 104013}, [\href{https://arxiv.org/abs/hep-th/9912259}{{\ttfamily hep-th/9912259}}].

\bibitem{Canate:2020btq}
P.~Ca{\~n}ate, D.~Magos and N.~Breton, \emph{{Nonlinear electrodynamics generalization of the rotating BTZ black hole}}, \href{http://dx.doi.org/10.1103/PhysRevD.101.064010}{\emph{Phys. Rev. D} {\bfseries 101} (2020) 064010}, [\href{https://arxiv.org/abs/2002.00890}{{\ttfamily 2002.00890}}].

\bibitem{Cataldo:1999wr}
M.~Cataldo and A.~Garcia, \emph{{Three dimensional black hole coupled to the Born-Infeld electrodynamics}}, \href{http://dx.doi.org/10.1016/S0370-2693(99)00441-4}{\emph{Phys. Lett. B} {\bfseries 456} (1999) 28--33}, [\href{https://arxiv.org/abs/hep-th/9903257}{{\ttfamily hep-th/9903257}}].

\bibitem{Myung:2008kd}
Y.~S. Myung, Y.-W. Kim and Y.-J. Park, \emph{{Thermodynamics of Einstein-Born-Infeld black holes in three dimensions}}, \href{http://dx.doi.org/10.1103/PhysRevD.78.044020}{\emph{Phys. Rev. D} {\bfseries 78} (2008) 044020}, [\href{https://arxiv.org/abs/0804.0301}{{\ttfamily 0804.0301}}].

\bibitem{Lemos:1994xp}
J.~P.~S. Lemos, \emph{{Cylindrical black hole in general relativity}}, \href{http://dx.doi.org/10.1016/0370-2693(95)00533-Q}{\emph{Phys. Lett. B} {\bfseries 353} (1995) 46--51}, [\href{https://arxiv.org/abs/gr-qc/9404041}{{\ttfamily gr-qc/9404041}}].

\bibitem{Lemos:1995cm}
J.~P.~S. Lemos and V.~T. Zanchin, \emph{{Rotating charged black string and three-dimensional black holes}}, \href{http://dx.doi.org/10.1103/PhysRevD.54.3840}{\emph{Phys. Rev. D} {\bfseries 54} (1996) 3840--3853}, [\href{https://arxiv.org/abs/hep-th/9511188}{{\ttfamily hep-th/9511188}}].

\end{thebibliography}\endgroup

\end{document}